\documentclass[aps,prx,twocolumn,superscriptaddress,longbibliography]{revtex4-1}
\usepackage{amsmath,amssymb,dsfont,graphicx}

\usepackage{subfigure}
\usepackage{mathrsfs}
\usepackage{amsthm}
\usepackage{hyperref}
\usepackage{xcolor}
\usepackage{bbold}
\usepackage{ulem} 

\usepackage{dcolumn}
\usepackage{bm}
\usepackage{float}

\newcommand{\revM}[1]{#1}

\hypersetup{colorlinks=true, linkcolor=black, citecolor=black, urlcolor=blue}
\allowdisplaybreaks

\begin{document}

\newcommand{\bra}[1]{\langle #1|}
\newcommand{\ket}[1]{| #1\rangle}
\newcommand{\tr}[1]{{\rm Tr}\left[ #1 \right]}
\newcommand{\av}[1]{\left\langle #1 \right\rangle}
\newcommand{\proj}[1]{\ket{#1}\bra{#1}}
\renewcommand{\k}{{\bf k}}
\newcommand{\x}{{\bf r}}
\newcommand{\bk}{\mathbf{k}}
\newcommand{\bq}{\mathbf{q}}

\title{Quantum-Zeno Fermi polaron \revM{in the strong dissipation limit}}

\author{Tomasz Wasak}
\affiliation{Max-Planck-Institut f\"ur Physik komplexer Systeme, N\"othnitzer~Str.~38, 01187 Dresden, Germany}

\author{Richard Schmidt}
\affiliation{Max-Planck-Institut f\"ur Quantenoptik, Hans-Kopfermann-Str.~1, 85748 Garching, Germany}

\author{Francesco Piazza}
\affiliation{Max-Planck-Institut f\"ur Physik komplexer Systeme, N\"othnitzer~Str.~38, 01187 Dresden, Germany}

\begin{abstract}
The interplay between measurement and quantum correlations in many-body systems can lead to novel types of collective phenomena which are not accessible in isolated
systems. In this work, we \revM{merge} the Zeno-paradigm of quantum
measurement theory with the concept of polarons in condensed-matter physics. The resulting quantum-Zeno Fermi-polaron is a quasi-particle which emerges for lossy
impurities interacting with a quantum-degenerate bath of fermions.
For loss rates of the order of the impurity-fermion binding energy the quasi-particle is short
lived. However, we show that in the \revM{strongly dissipative regime} of large loss rates a long-lived polaron branch re-emerges.
This quantum-Zeno Fermi-polaron originates from the nontrivial interplay between the Fermi-surface and the surface of the momentum region forbidden by the quantum Zeno projection. 
The situation we consider here is realized naturally for polaritonic impurities in charge-tuneable semiconductors and can be also implemented using dressed atomic states in ultracold gases.
\end{abstract}

\maketitle

\section{Introduction}

The effect of measurement on the time-evolution of a system is one of the most puzzling aspects of quantum dynamics \cite{breuer2002theory,wiseman2009quantum}, making it
drastically distinct from its classical counterpart. One paradigmatic example is the quantum Zeno effect \cite{misra_1977,itano_1990,PRESILLA199695}, whereby a frequent-enough
measurement of a quantum system results in a localisation of its state in Hilbert space \cite{facchi_2002}. Increasing the measurement rate from zero, the system dynamics
first gains a non-unitary character due to the loss of information to the measurement device. However, above a threshold-rate the trend is reversed: the dynamics tends
asymptotically to a unitary limit within a Hilbert space reduced by the region that is forbidden due the measurement process.

Recently, growing attention is being devoted to the effect of repeated measurements on the dynamics of quantum many-body systems
\cite{elliott_2015,mazzucchi_2016,dhar_2016,cao2018entanglement,li_2018,chan_2019,skinner_2019}. This has been motivated by experimental developments in controlling and
selectively coupling complex quantum systems to their environment, as demonstrated with superconducting qubits \cite{Katz_2006,vijay_2012,campagne_2016}, ultracold atoms
\cite{syassen2008strong,durr2009lieb,garcia2009dissipation,barontini_2013,labouvie_2016} and molecules \cite{yan2013observation,zhu2014suppressing}. The interplay between
measurement and intrinsic quantum correlations in many-body systems can lead to novel collective phenomena.  Recently, two instances of such collective phenomena
appearing in the quantum-Zeno limit have been predicted.  \revM{First, analyzing the entanglement entropy, it was found that by increasing the measurement rate of a
  {closed} quantum many-body system, a phase transition from a volume-law entangled phase to a quantum Zeno phase with area-law entanglement can take place
  \cite{li_2018,skinner_2019}}. Second, a study of one-dimensional fermionic models with \revM{a localized dissipation (of Dirac delta type)} revealed that the
quantum Zeno projection is enhanced in the vicinity of the Fermi surface due to the existence of gapless modes~\cite{froeml_2019}.

\revM{In this work, we consider a situation where a strong dissipation acts on a subspace of the Hilbert space of an interacting quantum system, with the following question in
mind: how are the many-body states affected by the emergence of a dissipation-induced constraint for the Hilbert space as the quantum-Zeno limit is approached? We focus
on the paradigm of mobile quantum impurity problems at the border between few- and many-body physics and extend it to the regime of gain and loss of impurity
particles. We demonstrate how the intrinsic correlation scales, governing the impurity problem, have a nontrivial interplay with the scale brought in by the
dissipation-induced constraint, leading to qualitative modification of the many-body quantum states.}

The concept of polarons plays a central role in condensed matter physics, developed to understand the motion of an electron moving in a dielectric crystal
\cite{landau1933bewegung,landau1948effective,frohlich1950properties,feynman1955slow}. The Fermi polaron, describing a mobile impurity interacting with a bath of Fermions
\cite{kopp_1995}, is particularly important for the quantitative description of the physics of imbalanced Fermi mixtures~\cite{chevy2006universal,combescot2008normal} and
sets the basis for our understanding of other strongly correlated systems~\cite{duine2005itinerant,cui2010stability,pietila2012pairing,massignan2014polarons}.  The
physics of Fermi polarons has become the center of significant interest due to recent experimental advances that allowed for its observation both with ultracold
atoms~\cite{schirotzek2009observation,kohstall2012metastability, koschorreck2012attractive} and more recently in charge-tunable monolayer semiconductors
\cite{sidler2017fermi,tan2019optical}.

So far, theoretical studies of the polaron paradigm are assuming a conserved number of impurities in a closed system.  However, in addition to the fundamental interest in
exploring the effect of strong measurement on quantum many-body systems in general, the recent experiments in charge-tunable monolayer semiconductors open a new frontier
to study polarons in the presence of impurity loss and gain, as the corresponding rates can become comparable to the timescales of unitary dynamics
\cite{sidler2017fermi}.  In this work we develop a theoretical approach to describe such quantum driven-dissipative impurity dynamics.  Using a diagrammatic, real-time
Keldysh approach we show that, while for intermediate loss rates the polaron becomes expectedly short-lived, in the Quantum Zeno limit of large loss rates a long-lived
polaron branch re-emerges. This Quantum-Zeno Fermi polaron is not simply a copy of the original polaron in absence of losses, but originates from the nontrivial interplay
between the Fermi-surface and the surface of the region forbidden by the quantum Zeno-projection. 
\revM{This indeed leads to qualitative modifications of the dispersion of the impurity-fermion bound states as well as of the scattering continuum.}

This complex interplay underlying the formation of the Quantum-Zeno Fermi-polaron requires a loss profile that can single-out a non-trivial region in momentum space.
This can be achieved by mixing the impurity with an additional lossy degree of freedom of much smaller mass. This is naturally the case for exciton-polariton impurities
in charge-tunable monolayer semiconductors. It can also be engineered for ultracold atoms by coupling a stable electronic level to a short-lived level, and additionally
using level-depending trapping potentials.

\revM{This manuscript is structured as follows.
In Section~\ref{model-approach}, we introduce a model describing the driven-dissipative Fermi-polaron. 
In Section~\ref{sec-res}, we discuss the results for the spectral response of the system. We show the emergence of the quantum-Zeno Fermi-polaron for large loss rates and
illustrate the interplay between the surface of the Zeno-forbidden region and the Fermi surface.
In Section \ref{sec:realization}, we discuss in detail the concrete implementation of the Fermi-polaron subject {to strong dissipation}.
Finally, in Section~\ref{sec-conc}, we offer some concluding remarks and an outlook.
Technical details are presented in Appendices, where, in particular, 
we introduce a Keldysh diagrammatic approach to non-equilibrium Green's functions.
}


\revM{\section{Driven-dissipative impurity in a Fermi bath}
\label{model-approach}
\label{diss-drive}
\label{noneqGF}
}

We consider a system composed of a bath of fermions interacting with a vanishingly small density of impurity particles.
The Hamiltonian of the system consists of three parts:
\begin{equation}
  \label{htotal}
  \hat H = \hat H_c + \hat H_f + \hat H_\mathrm{int},
\end{equation}
where $\hat H_c = \sum_\bk \varepsilon_c(\bk)\hat c_\bk^\dagger \hat c_\bk$ and $\hat H_f = \sum_\bk \varepsilon_f(\bk) \hat f_\bk^\dagger \hat f_\bk$ are free-particle
Hamiltonians with kinetic energies $\varepsilon_c(\bk) = \bk^2/2m_c$ and $\varepsilon_f(\bk)=\bk^2/2m_f$ of impurities and fermions, respectively.  We assume that the
interaction is a contact potential characterized by a coupling strength $U$, i.e.,
\begin{align}
  \label{eq:Hint}
  \hat H_\mathrm{int} = U \int\!\! d\x\, \hat c^\dagger(\x) \hat c(\x) \hat f^\dagger(\x) \hat f(\x).
\end{align}
\revM{In this manuscript,} all concrete calculations will be given for the two-dimensional case which has been realized in ultracold atoms~\cite{koschorreck2012attractive,thomas_polaron_2012} and
two-dimensional semiconductors~\cite{sidler2017fermi,tan2019optical}.

Driven-dissipative impurities which are localized \revM{and immobile} in a one-dimensional system have been recently studied in Refs.~\cite{Tonielli_2019,Berdanier_2019}.
{Here we consider mobile impurities which} are subject to loss and thus require to be repumped into the system. 
To model the action of dissipation and pump, we employ the master equation
\begin{equation}
\label{lind}
  \partial_t \hat \varrho(t) = -i [\hat H, \hat{\varrho}(t)] + \mathcal{L}_d \hat{\varrho}(t).
\end{equation}
The dissipative Lindblad operator $\mathcal{L}_d$, acting on the density matrix of the whole system $\hat \varrho(t)$, splits into two parts, i.e., $\mathcal{L}_d
\hat\varrho = \sum_\bk \{ \gamma(\bk) D[\hat{c}_\bk] + \Omega(\bk) P[\hat c_\bk] \} \hat \varrho$, which describe loss of particles with rate $\gamma(\bk)$ and their
reinjection with rate $\Omega(\bk)$. The Markov form of the master equation \eqref{lind} is valid as long as the reservoir of impurity-particles has a correlation time
much shorter than all other timescales characterizing the system \cite{breuer2002theory}. \revM{We model here the dissipation strength by a  momentum dependent profile, 
as it may be generated by a coupling of an impurity to a lossy degree of freedom, for example excitons to microcavity photons or ground state atoms to short-lived excited states.}

We consider loss and injection of single impurities. These two incoherent processes are expressed by the super-operators $D[\hat c] \hat{\varrho} \equiv \hat c^\dagger
\hat{\varrho} \hat c - \frac12 \big\{\hat c^\dagger\hat c , \hat{\varrho} \big\}$ and $P[\hat c] \equiv D[\hat c] + D[\hat c^\dagger]$. The action of each super-operator
generates two terms. The anti-commutator term corresponds to a non-Hermitian Hamiltonian evolution describing the damping/growth of the number of impurities, while the
other term describes the noise which necessarily accompanies dissipation. In a stochastic quantum trajectory picture \cite{breuer2002theory}, the noise-term generates
projections of the wave function at random times induced by the measurement, which, in our case, corresponds to the detection of an impurity particle exiting or entering
the system. In this picture, both the measurement and the damping (or growth) rate are set by $\gamma(\bk)$ (or $\Omega(\bk)$).  For the present study, the specific form
of the pump term $P$ is irrelevant since the excitation spectrum of the system does not depend on it as long as the average density of impurities is small.

In the limit of vanishing dissipation and drive, our model corresponds to the Fermi-polaron problem \revM{treated as a closed system}. 
One approach to describe the polaronic states of an impurity coupled
to a bath of fermions in this closed system, is to employ a variational approach based on the so-called Chevy
ansatz \revM{for the wave function}~\cite{chevy2006universal,combescot2007normal,zollner2011polarons, parish2011polaron,massignan2014polarons,kroiss2014diagrammatic}. In this ansatz the impurity
motional degree of freedom is entangled with a single particle-hole excitation of the Fermi sea. The variation with respect to the expansion coefficients yields an
equation that describes the properties of the lowest lying polaron state, the so-called attractive polaron. Further variational methods, including higher-order
particle-hole excitations, have also been developed to improve the Chevy expansion~\cite{combescot2008normal}. The variational approach~\cite{ngampruetikorn2012repulsive}
and a related non-self-consistent $T$-matrix approach~\cite{schmidt2012fermi} were used to capture the properties of an excited state of the polaron, known as the
repulsive polaron. All these theoretical studies rely on the Hamiltonian description of the system and, therefore, are not directly applicable to the case of
driven-dissipative impurities.

\revM{In this work, we develop a diagrammatic approach based on non-equilibrium Green's functions (GFs) which extends the non-self-consistent $T$-matrix approach to
driven-dissipative systems.}  We employ a path-integral formulation of our problem on the Keldysh real-time contour which allows for a straightforward inclusion of the
Lindblad terms of the master equation~\cite{kamenev2011field, sieberer2016keldysh}. Within this formalism, diagrammatic approaches are applicable in a way analogous to
the equilibrium Matsubara formalism. In the following, we analyze the properties of the system in its steady state, where the GFs are time-translation invariant. 
\revM{The details of the calculations can be found in Appendix~\ref{app-GFs}.}

\revM{In what follows, we will focus on the impurity limit where the number of impurities in the system is vanishingly small
and the \revM{excitation spectrum of the system } 
will not depend on the pump $\Omega(\bk)$ or the quantum statistics of impurities.}

%
%
\begin{figure*}[tb]
  \centering \includegraphics[width=1.0\textwidth]{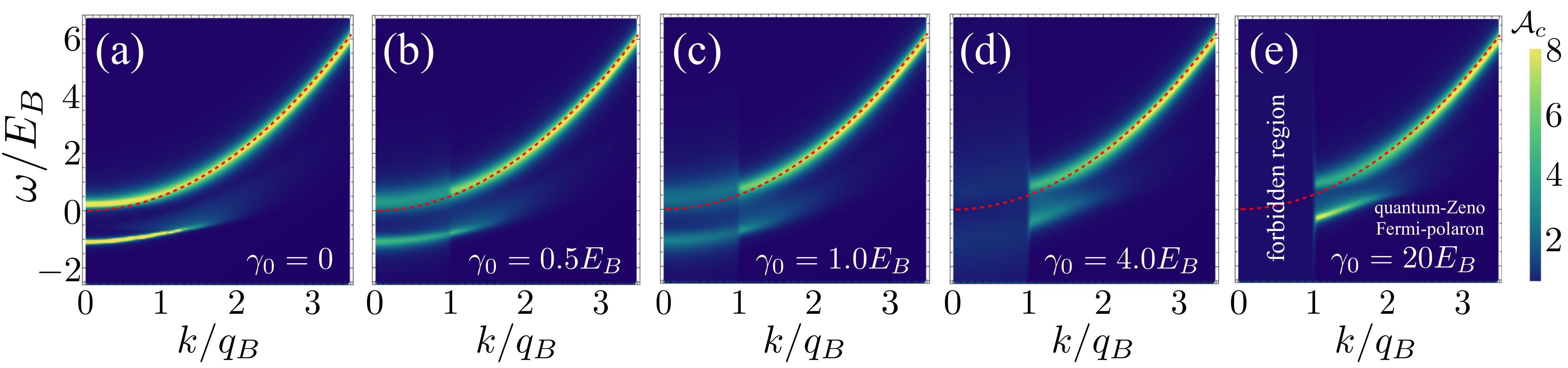}
  \caption{The impurity spectral function $\mathcal{A}_c(k,\omega)$ (in the units of $E_B$) for a 2D system with the mass balance $m_c = m_f$. In the panels from (a)--(e)
    the strength of the loss is increased: $\gamma_0/E_B =$ 0, 0.5, 1.0, 4.0, 20.  The energy is expressed in the units of the impurity-fermion binding energy $E_B$, and
    momenta are given in $q_B = \sqrt{m_f E_B}$.  The Fermi energy $\epsilon_F = k_F^2/2m_f$ is set by the Fermi wavevector $k_F = 0.7 q_B$. The color coding of the
    values for $\mathcal{A}_c$ is on the right.  The cutoff momentum of the loss profile is at $k_\gamma = q_B$. The dashed red curve is the free dispersion relation $\omega =
    \varepsilon_c(k) \equiv k^2/2m_c$.}
\label{fig:Ax}
\end{figure*}

%
%
\begin{figure*}[!htb]
  \centering
  \includegraphics[width=\textwidth]{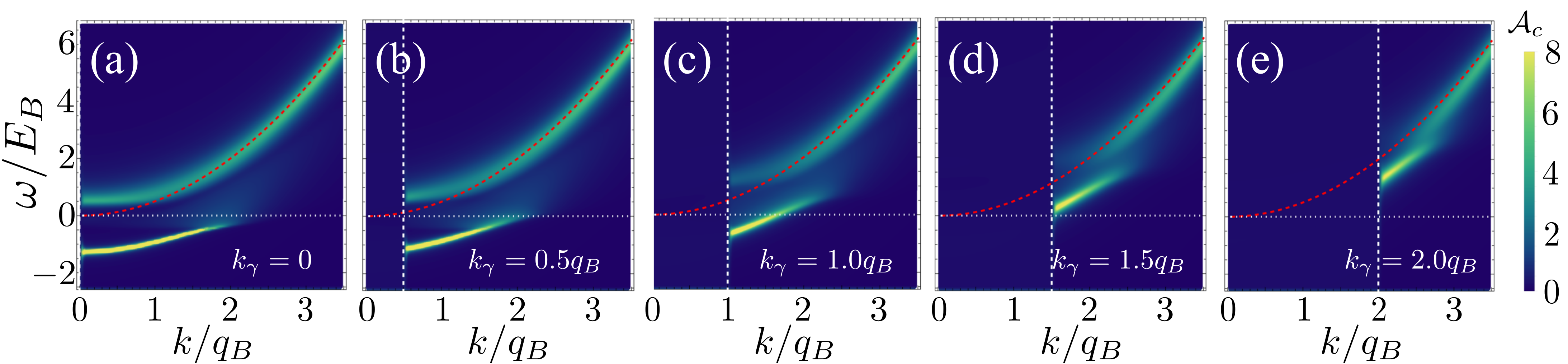}
  \caption{Impurity spectral function $\mathcal{A}_c(\bk,\omega)$ (in units of $E_B^{-1}$) as a function of $|\bk|$ and $\omega$ (in units of $q_B$ and $E_B$, respectively) 
    for different sizes of the
    dissipative subspace: $k_\gamma/q_B = 0, 0.5, 1.0, 1.5, 2.0$ [from (a) to (e)]. The Fermi energy $\epsilon_F = k_F^2/2m_f$ is set by $k_F = q_B$, and $\gamma_0 = 20\,
    E_B$. The dotted horizontal line depicts $\omega=0$. The dashed vertical line [in the panels (b)--(e)] displays $k_\gamma$. The dashed red parabolic curve shows the
    dispersion relation $\omega = k^2/2m_c$ of the free impurity.}
  \label{fig:Axs}
\end{figure*}

%
%
\begin{figure}[tb]
  \centering \includegraphics[width=1\columnwidth]{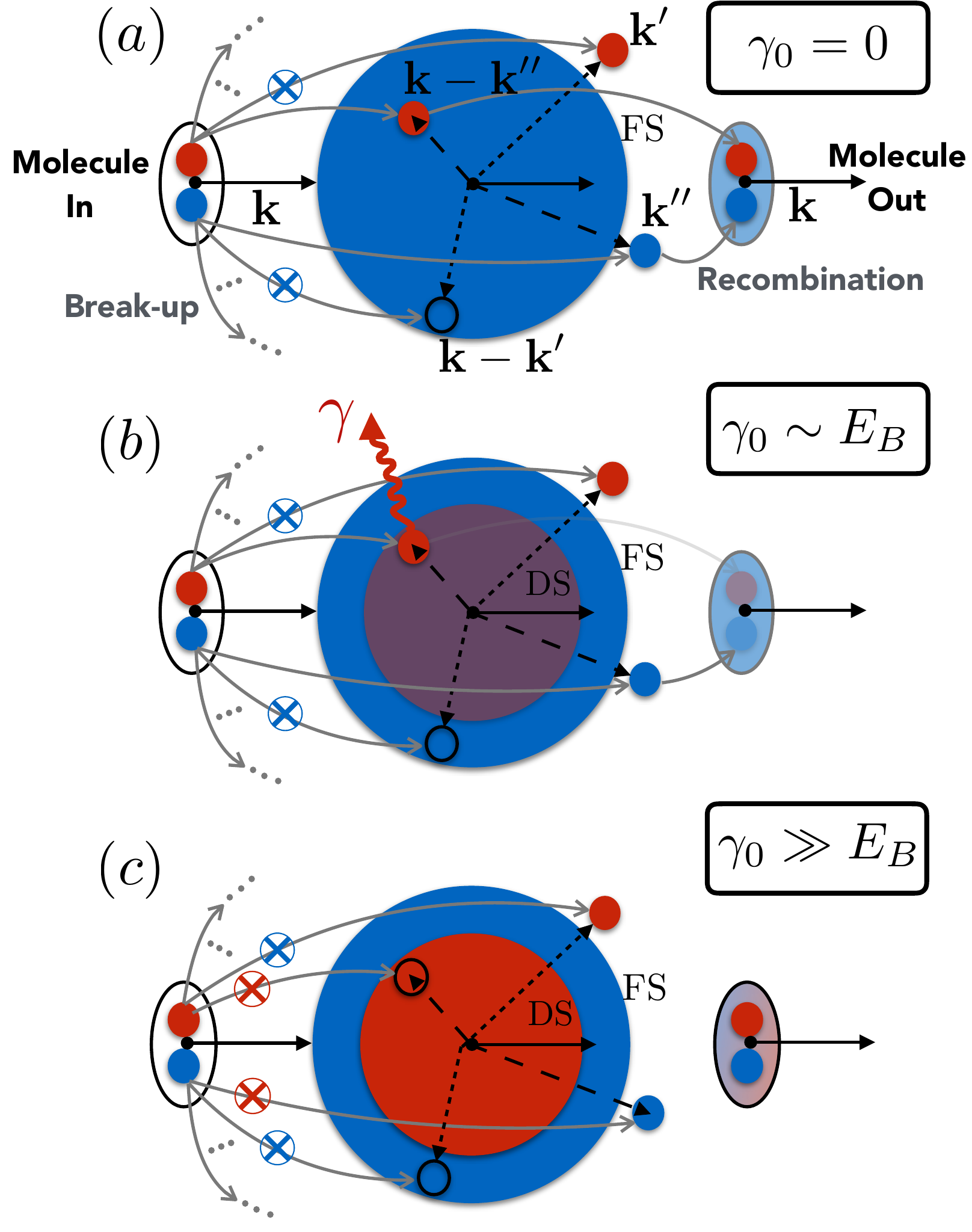}
  \caption{The polaron is formed because the impurity (red circle) forms a molecular state with a fermion (blue circle). Crucially, the dynamics of molecules is in turn
    generated by virtual break-up processes followed by recombination, as illustrated here pictorially to complement the diagrammatic representation of
    Fig.~\ref{fig:diag_xD}b and the formal expression of Eq.~\eqref{SED} for the molecular self-energy. A molecule with momentum $\mathbf{k}$ temporarily breaks up into
    an impurity with momentum $\mathbf{k}'$ and a fermion with momentum $\bk-\bk'$. The total effect is obtained by summing over all processes involving different
    momenta, as graphically indicated by the dots.  In (a) the dissipationless case is considered, where the Pauli principle forbids processes creating a fermion within
    the Fermi surface (FS).  In panel (b), moderate dissipation at a rate $\gamma_0$ comparable with the impurity-fermion binding energy $E_B$ is added within the DS,
    which effectively makes the molecular state dissipative as well.  Finally, in (c) the strongly dissipative regime is considered, where the quantum-Zeno effect forbids
    the creation of an impurity within the DS.  }\label{fig:FSDS}
\end{figure}

\section{Fermi-polaron with a lossy region in impurity-momentum space}
\label{sec-res}

\subsection{Impurity response function: quantum-Zeno Fermi-polaron}
\label{sec-imp-SF}

As we shall see, the emergence of a quantum-Zeno Fermi-polaron relies on impurity losses which single out at least one subset of momentum space. Such a subset we call the
dissipative subspace (DS). In this section, we consider the simplest case of a loss function $\gamma(\bk)$ that is characterized by a single scale given by a cutoff
momentum~$k_\gamma$, and we assume that $\gamma(\bk) = \gamma_0$ if $|\bk| < k_\gamma$ and is vanishing otherwise.  Our approach is valid for any value of the loss rate
$\gamma_0$.  However, we will be interested in the case of large loss rates: $\gamma_0\gg E_B$ with $E_B$ the impurity-fermion binding energy. 
\revM{This leads to effective blocking of transitions between states involving the DS, which can be interpreted in terms of the quantum Zeno effect.}  
The exact form of the loss profile is not crucial for the physics described in what follows, and it is sufficient that the loss rate decays fast enough over the
momentum scale $q_B = \sqrt{m_f E_B}$. We will demonstrate this by considering experimentally realistic cases in Sec.~\ref{sec:realization}.
\revM{Finally, we remark that the coupling strength $U$ is eliminated from the equations by finding the impurity-fermion binding energy $E_B$ (based on the poles of the vacuum
$T$-matrix) in the limit of vanishing fermion density and with dissipation strength set to zero.}

\revM{The spectral response of the system is encoded in the retarded impurity GF, which is given by
\begin{equation}
  \label{eq:impurity_GR_dressed}
  G_c^R(\bk,\omega) = \frac{1}{\omega - \varepsilon_c(\bk) + i \frac{\gamma(\bk)}{2}  - \Sigma^R_c(p) + i0},
\end{equation}
where the retarded self-energy $\Sigma^R_c(p)$ of the impurity contains information about the interaction with the surrounding bath of fermions; here and below the four-momentum
$p \equiv (\bk,\omega)$. The spectral function is then given by the following formula:}
\begin{equation}
  \label{defAx}
  \mathcal{A}_c(\bk,\omega) = - 2 \mathrm{Im}[G_c^R(\bk,\omega)].
\end{equation}
The spectral function satisfies the exact sum rule $\int\!\mathcal{A}_c(p) d\omega/(2\pi)= 1$.  It exhibits resonant peaks in correspondence to quasi-particle excitations
with a width related to their inverse lifetime.  In the absence of interactions with the fermions, the impurity spectral function $\mathcal{A}_{0,c} = - 2
\mathrm{Im}[G_{0,c}^R]$ has resonance at its bare dispersion $\omega = \varepsilon_{c}(\bk)$, with a width determined by $\gamma(\bk)$.  In presence of interactions,
resonances appear at the renormalized frequencies $\omega = \varepsilon_c(\bk) - \mathrm{Re}[\Sigma_c^R(\bk,\omega)]$. \revM{In this case, the inverse lifetime is set by the
imaginary part of the pole position on the complex plane: $\mathrm{Im}[\omega(\mathbf{k})]$ satisfying $[G_c^R(\bk,\omega(\mathbf{k}))]^{-1}=0$. For sufficiently large
inverse-lifetime the latter is approximately given by $\mathrm{Im}\{[G_c^R(\bk,\omega)]^{-1}\} =\mathrm{Im}[-\Sigma_c^R(\bk,\omega) + i \gamma(\bk)/2]$.}

An example of the impurity spectral function for gradually increasing $\gamma_0$ is presented in Fig.~\ref{fig:Ax} for the mass balanced case. For small $\gamma_0$, two
quasi-particle resonances are visible. The lower one in energy is the so-called attractive polaron. It corresponds to a state where the impurity attracts a cloud of
surrounding fermions. The attractive polaron is long-lived (the width of the peak vanishing for $|\bk|\to 0$) and ceases to exist for large momenta, when the resonance
enters the molecule-hole continuum. The latter appears as blurred region between the two polaron resonances.  The second resonance is higher in energy and corresponds to
a repulsive polaron state where the surrounding cloud is repelled by the impurity. For low-enough losses and small enough momenta, the repulsive polaron has a shorter
lifetime than the attractive polaron.

It is important to note that at any finite value of $\gamma_0$, while the losses affect bare impurity only within the DS defined by $|\bk|<k_\gamma$, they do affect the
polaron also {\it outside} the DS. The reason for this is that the polarons are impurities dressed by the interactions with the surrounding fermions and therefore involve
a superposition of all the allowed momentum states. This statement is particularly explicit in the simple variational ansatz introduced at the end of \revM{Appendix}
\ref{sec-dyson}. 
\revM{Impurity losses thus affect the polarons in a non-trivial manner via the coupling between the impurity and the molecular state. The latter is very sensitive to dissipation.}
We note that for the value of $k_\gamma$ used in Fig.~\ref{fig:Ax} and for momenta outside the DS, the repulsive polaron remains largely unaffected by the losses
(see also Fig.~\ref{fig:Axs}).  This is due to small coupling of the impurity with the bound state at high energies, where the repulsive polaron recovers the impurity
dispersion relation. The effect of losses outside the DS is most clearly visible in panels (c) and (d) in Fig.~\ref{fig:Ax}, where, for $\gamma_0 \sim E_B$, the
attractive-polaron peak is almost fully washed out.

Quite remarkably, for even larger losses $\gamma_0 \gg E_B$, a resonance reappears below the repulsive polaron just outside the DS: $k>k_\gamma$. This resonance has a
width which does not depend anymore on the loss rate $\gamma_0$ and is rather given only by the coupling to the molecule-hole continuum, as would be the case without
losses.  At the same time, the DS does not contain any spectral weight and is fully forbidden for the excitations.  This phenomenology can be understood within the
paradigm of the quantum-Zeno effect. Indeed, when the impurity occupies a state with $\bk\in\mathrm{DS}$ it is lost immediately, which is equivalent to a frequent
measurement performed within the DS. As a result, the dynamics of the particle is confined to the orthogonal subspace, i.e. $\bk\notin\mathrm{DS}$.  \revM{We note that
  for infinite dissipation strength, the equation for the pole of the impurity Green's function can be reproduced by a modification of the Chevy polaron ansatz for the
  wave function in which the momenta from DS are excluded.}

The loss rate $\gamma_0$ does not appear explicitly anymore as the dynamics becomes unitary again in the quantum Zeno limit $\gamma_0 \gg E_B$ (see also Fig.~\ref{fig:S}
in the next section). Still, the size of the DS $k_\gamma$ remains as further scale which plays a central role in the problem, as does the Fermi wavevector~$k_F$.  One
effect of the non-trivial interplay between these scales is demonstrated in Fig.~\ref{fig:Axs}, where we show how the polaron dispersion depends on the size of the DS.
By increasing $k_\gamma$ the attractive-polaron dispersion gradually changes from quadratic to linear. Moreover, the branch remains well defined up to larger values of
$k$ compared to the lossless case and is also blueshifted. This is evident from Fig.~\ref{fig:Axs}(d) where the resonance appears also for $\omega>0$. In addition, the
spectral weight of the repulsive polaron is transferred to the attractive branch. By increasing $k_\gamma$ the attractive polaron peak becomes broader due to the
entrance into the molecule-hole continuum.

\subsection{Molecular response function: quantum-Zeno scattering}
\label{sec-mol-SF}

In the interaction between an impurity and a fermion, the formation of paired (or molecular) states plays a crucial role. Our approach based on an auxiliary molecular
field allows to rewrite the impurity self-energy in the following way, \revM{see Appendix~\ref{app-GFs}}:
\begin{equation}
  \Sigma_c^R(\bk,\omega) = \frac{1}{V}\!\! \sum_{\bk' \in \mathrm{FS}} G_\Delta^R\big(\bk+\bk',\omega +\varepsilon_f(\bk')\big),  
\end{equation}
from which we see that the renormalization of the impurity is fully determined by the molecular GF. The molecular GF is at the same time directly related to the
$T$-matrix describing the scattering between the impurity and a fermion. Specifically, we will define $T(\bk,\omega) = G_\Delta^R(\bk,\omega)$, where $\bk$ is the
momentum of the molecule and $\omega$ is its energy. \revM{In Appendix~\ref{sec-dyson} we calculate the molecular-self energy in the impurity limit, which takes the form:
\begin{equation}
\label{SED2}
  \Sigma_\Delta^R(\bk,\omega) \!\!=\!\! 
  \frac{1}{V} \!\! \sum_{\bk' \notin \mathrm{FS}} \frac{1}{\omega\! -\! \varepsilon_c(\bk - \bk') \!-\! \varepsilon_f(\bk') \!+\! i \frac{\gamma(\bk - \bk')}{2}\! +\! i0}.
\end{equation}
This object then determines the molecular GF, which reads:
\begin{equation}
  \label{eq:molecule_gf}
  G_\Delta^R(p) = \frac{1}{- \nu^2 - \Sigma_\Delta^R(p)+i0},
\end{equation}
where $\nu^2 = -1/U$.}

In the case of a single fermion interacting with a lossless impurity, which is recovered by sending both $\epsilon_F$ and $\gamma_0\to0$, we obtain (in two-spatial
dimensions) the usual form of the vacuum $T$-matrix:
\begin{equation}
  \label{T0}
  T^{-1}(\bk,\omega) = - \frac{m_r}{2\pi}\bigg[ \ln\bigg(\frac{E+i0}{E_B}\bigg) -i \pi\bigg] \equiv T_0^{-1}(E),
\end{equation}
which depends only on the kinetic energy of the relative motion of the particles $E=\omega - k^2/2M$ with $M=m_c+m_f$ the total mass; the index ``0'' in $T_0$ refers
to the vacuum case without losses. The $T$-matrix has a pole at $E=-E_B$ signalling the presence of a bound state at that energy.

For $\epsilon_F>0$, the scattering properties are modified due to the presence of the fermionic medium.  Here, an important role is played by the Pauli exclusion
principle, which results in the constraint $\bk'\notin\mathrm{FS}$ on the summation variable in $\Sigma_\Delta^R$ (see Eq.~\eqref{SED2}). This restricts the number of
scattering processes which modify the molecule-propagation. As already pointed out, these processes are the only source of the dynamics of the molecular state, since the
only frequency-momentum dependence of the molecule GF from Eq.~\eqref{eq:molecule_gf} is induced by the self-energy $\Sigma_\Delta^R$. \revM{As shown pictorially in
Fig.~\ref{fig:FSDS}, and diagrammatically in Fig.~\ref{fig:diag_xD}b}, the molecule dynamics receives a contribution from all energy-and-momentum-conserving processes
where a molecular state temporarily breaks up into an impurity plus a fermion. In absence of losses, corresponding to panel (a) in Fig.~\ref{fig:FSDS}, all processes
creating a fermion within the FS are forbidden by Pauli blocking.  This case has been already studied in the literature for the mass-balanced case
\cite{randeria1989bound,engelbrecht1990new,engelbrecht1992low}, and the self-energy can be computed analytically~\cite{schmidt2012fermi}.  The result can be generalized
to the mass-imbalanced case~\cite{efimkin_imbalanced_2017}, but we do not present it here.

We now turn to the situation where impurity losses are present in the DS.  The relevant processes in this case are shown in Fig.~\ref{fig:FSDS}b: whenever the molecule
breaks up creating an impurity within the DS, there is a finite probability for the impurity to be lost so that the molecule cannot be rebuilt. Considering the sum over
all processes, this induces dissipation for the dressed molecular states.  From Eq.~\eqref{SED2} we see that for momenta $\bk'$ satisfying $\bk-\bk'\in\mathrm{DS}$, the
contribution to $\Sigma_\Delta^R$ is negligible if the loss rate is very large: $\gamma_0\gg E_B$. If the number of fermions is vanishing, performing a shift in $\bk'$ we
can rewrite the sum as:
\begin{equation}
  \label{SED-zeno}
  \Sigma_\Delta^R(\bk,\omega)\bigg |_{\epsilon_F=0=\frac{1}{\gamma_0}} \!\!\!=\! 
  \frac{1}{V} \!\! \sum_{\bk'' \notin \mathrm{DS}} \frac{1}{\omega \!-\! \varepsilon_f(\bk \!-\! \bk'') \!-\! \varepsilon_c(\bk'')  \!+\! i0}.
\end{equation}
Comparing Eq.~\eqref{SED2} for $\gamma=0$ with Eq.~\eqref{SED-zeno}, we see that the dissipative subspace in the limit of large losses plays a role analogous to a Fermi
surface, as illustrated in Fig.~\ref{fig:FSDS}c. In other words, at the level of scattering physics the forbidden DS in momentum space results from an effective ``Pauli
blocking'' emerging in the quantum-Zeno limit~\cite{syassen2008strong}.

The analogy between the forbidden DS and the FS is however not complete. For example, one can excite fermions out of the FS but not the DS. As for the analogous case of a
FS, the sum in Eq.~\eqref{SED-zeno} can be evaluated analytically in presence of the forbidden DS, leading to
\begin{align}
  T(\bk,\omega) &\bigg |_{\epsilon_F=0=\frac{1}{\gamma_0}}\!\!\! = T_0\bigg[ \frac{z}{2} - \bigg(\frac12 - \frac{m_r}{m_f}\bigg)\varepsilon_f(\bk) + \nonumber\\ 
    & \pm \frac12 \sqrt{\big[z - \varepsilon_f(\bk)\big]^2 - 4 \varepsilon_f(\bk) \frac{k_\gamma^2}{2 m_f}}\bigg], 
  \label{SED-zeno2}
\end{align}
where $z = \omega - \frac{k_\gamma^2}{2 m_r} + i0$, and $\pm = \mathrm{sign}[\mathrm{Re}(z- \varepsilon_f(\bk))]$. Interestingly, in this case the $T$-matrix can be
expressed in terms of the vacuum two-body $T_0$ matrix. This is a property of the 2D geometry~\cite{schmidt2012fermi} and does not occur in other dimensions.

In order for the quantum-Zeno Fermi-polaron to appear, the existence of both a FS and a forbidden (or at least strongly dissipative) DS is required. In this case the molecular self-energy can be rewritten as
\begin{equation}
  \label{SED-zeno3}
  \Sigma_\Delta^R(\bk,\omega) \bigg |_{\frac{E_B}{\gamma_0}\simeq 0} \!\!\!\simeq\! 
  \frac{1}{V} \!\! \!\!\sum_{\substack{\bk' \notin \mathrm{FS}\\\bk''\notin \mathrm{DS}}}{}\!\!\, \frac{{\delta_{\mathbf{k''+k',k}}}}{\omega \!-\! \varepsilon_c(\bk'') 
    \!-\! \varepsilon_f(\bk') + i0}, 
\end{equation}
where the sum over all processes is subject to {\it two} constraints.

An important feature of the quantum-Zeno limit is that the emergence of a forbidden subspace coincides with the reemergence of unitary dynamics. This becomes evident when
evaluating the scattering S matrix. In closed-system dynamics, the $S$ matrix conserves the total probability as expressed by the unitarity condition $ S^\dagger S =
\mathbb{1}$, where $\mathbb{1}$ is the operator identity. In the presence of dissipation, we have instead a decrease of the probability for some of the scattering
channels and consequently $|e^{i\delta}|<1$, where the phase shift $\delta(E)$ characterizes the eigenvalue $s=e^{i\delta}$ of the scattering $S$ matrix at a collision
energy $E=E_\mathrm{rel}=k^2/2m_r$. In the quantum-Zeno limit, we thus expect $|s|\to 1$ for $\gamma_0\to\infty$.

To proceed further, we relate $s$ to the $T$-matrix. For two-body collisions in 2D without dissipation, the on-shell $T$-matrix is related to the $S$-matrix
by~\cite{randeria1990superconductivity,engelbrecht1992low}:
\begin{equation}
  \label{s0def}
  s_0(\bk,\omega) = 1 + \frac{i t_0(\bk,\omega)}{2},
\end{equation} 
where for our case of $s$-wave collisions, $t_0 = - 2 m_r T_0$; $\bk$ is the momentum of the center of mass and $\omega = E_\mathrm{rel} + k^2/2M$ is the total energy,
i.e., the sum of the kinetic energy of the center of mass and the collision energy of relative motion $E_\mathrm{rel}$. This relation implies $s_0 =
e^{i\delta_\mathrm{0}}$ and in particular $|s_0|^2=1$ for the scattering states with $E_\mathrm{rel} > 0$, as required by unitarity.

In our open system, we generalize the relation Eq.~\eqref{s0def} to the case where one of the two scattering partners is subject to losses (this is different, but related
to the case of inelastic collisions \cite{braaten2016open}):
\begin{equation}
  \label{sdef}
  s(\bk,\omega) = 1 + \frac{i t(\bk,\omega)}{2},
\end{equation} 
where $t = - 2 m_r T$ and $\omega = k^2/2M + E_\mathrm{rel}$.  This leads to $|s|=1$ for collisions in vacuum or in medium (without dissipation) while $|s|<1$ for nonzero
loss rate $\gamma_0$. These relations are satisfied provided that in all these cases the scattering states exist, that is, the states are in the continuum
$E_\mathrm{rel}>0$ and the particles can move away to infinity (we are not considering here cases where scattering states might not be well defined due to the form of the
scattering potential itself).  The second condition means that the momentum of the initial or final state should not be subject to loss from the DS. For an impurity
moving with momentum $\bk_c = \frac{m_c}{M}\bk + \bk_\mathrm{rel}$, where $\bk$ is the total momentum of the impurity-fermion system and $\bk_\mathrm{rel}$ is the
momentum of the relative motion, the condition $\bk_c\notin \mathrm{DS}$ implies that the particles can escape after the collision for all directions of
$\bk_\mathrm{rel}$, if
\begin{equation}
  \label{Smat-th}
  \bigg(\sqrt{2m_r E_\mathrm{rel}} - \frac{m_c}{M} k\bigg)^2 > k_\gamma^2.
\end{equation}
In order to simplify the analysis below, we set the threshold value to $E_\mathrm{rel}^\mathrm{th} \equiv \omega_\mathrm{th} - k^2/2M = \frac{[(1 - \frac{m_r}{m_c}) k +
    k_\gamma]^2}{2 m_r}$. We note that, while for for all collision energies $E_\mathrm{rel}> E_\mathrm{rel}^\mathrm{th}$ the condition Eq.~\eqref{Smat-th} is satisfied,
there might also be conditions for which it is satisfied in a window of energies $0 < E_\mathrm{rel} < E_\mathrm{rel}^\mathrm{th}$.

\begin{figure}[htp]
  \centering
  \includegraphics[width=0.9\columnwidth]{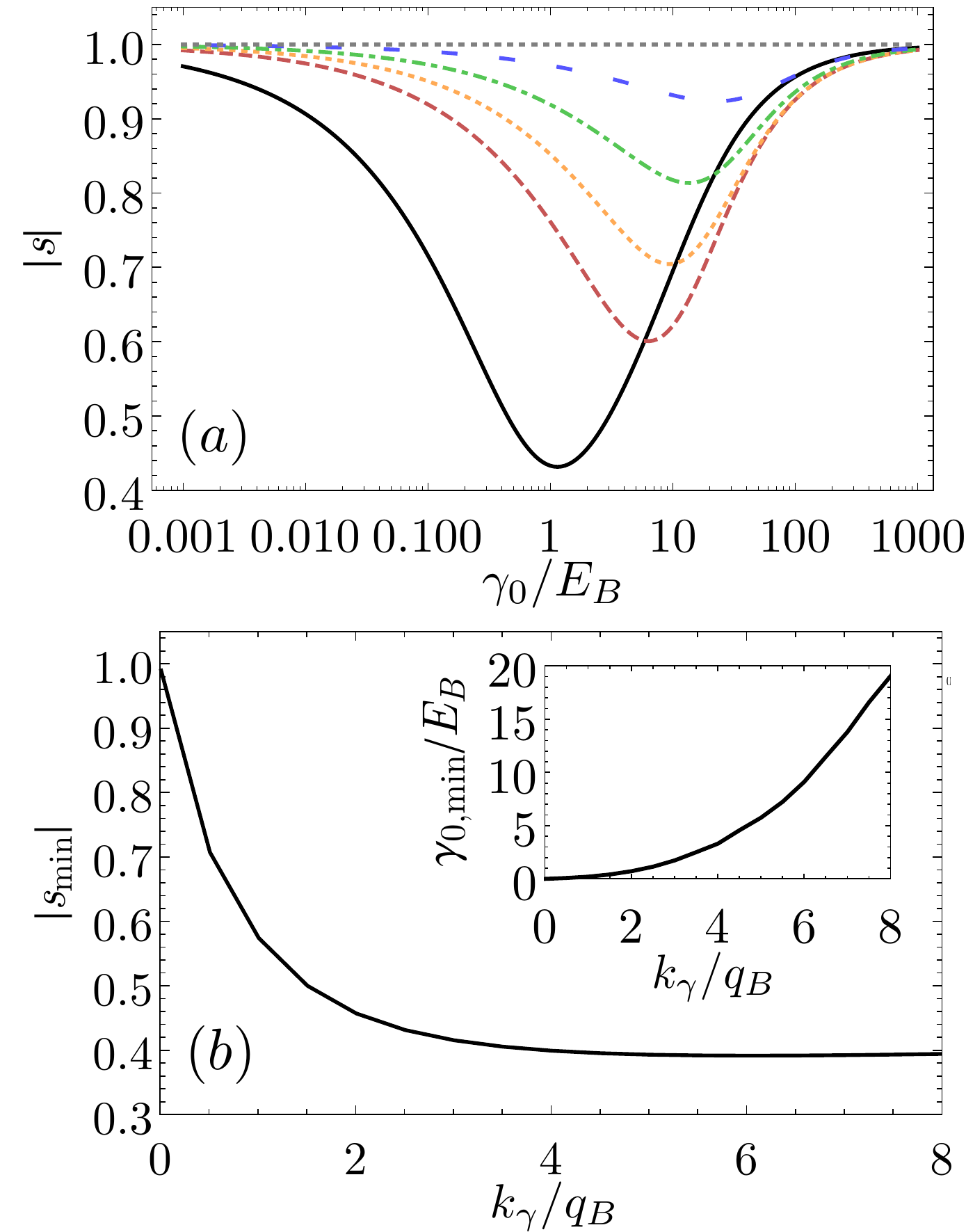}
  \caption{ 
    (a) The $S$-matrix: $|s(\bk,\omega)|=|e^{i\delta(\bk,\omega)}|$ as a function of the loss rate $\gamma_0$ (in units of $E_B$, computed at threshold
    $\omega=\omega_{\rm th}$ for $k/q_B = $ 0.05 (solid black), 0.5 (dashed red), 1 (orange dotted), 2 (dotted-dashed green), 5 (double-dotted-dashed blue); here
    $k_\gamma = 2.5 q_B$. The dotted gray horizontal line is for infinitely fast losses, see Eq.~\eqref{SED-zeno2}.
    (b) \revM{The minimum of $|s_\mathrm{min}| \equiv \min_{\gamma_0}|s(\gamma_0)|$ as a function of $k_\gamma$ calculated for $k = 0.05 q_B$.}
    The inset of panel (b) shows the value of $\gamma_0$ for which $|s|$ takes the minimum value, denoted by $\gamma_{0,\mathrm{min}}$. In both panels, $m_c=m_f$.}
\label{fig:S}
\end{figure}

\revM{
In Fig.~\ref{fig:S}a we show $|s|$ computed for the energy $E^\mathrm{th}_\mathrm{rel}$ as a function of the loss rate $\gamma_0$ for various total momenta
of the particles $k$ ranging from 0.05 to $5.0q_B$. We observe that for small values of $\gamma_0 \ll E_B$, the low momenta $k\ll k_F$ are affected by the loss, reducing
the value of $|s|$ below unity. Increasing the loss rate leads to a further decrease of $|s|$, as the scattering becomes dominated by the presence of DS. For low momenta, 
the deviation from unitarity becomes maximal when $\gamma_0$ is on the order of $E_B$. For still higher $\gamma_0$, the value of $|s|$ increases towards unitarity, as the 
contributions to $s$ from DS become negligible. In the limit of very large $\gamma_0$ we observe $|s|\to 1$. The deviations from unitarity are decreasing
with increasing the scattering momentum.

We observe that for a fixed value of $k$, the threshold value of $\gamma_0$ (called $\gamma_{0,\mathrm{min}}$), beyond which $|s|$ starts to increase (called
$|s_\mathrm{min}|$), is on the order of $E_B$. In Fig.~\ref{fig:S}b, we display the maximum violation of unitarity (main panel) and the threshold value $\gamma_0$ (inset), as a
function of the extent of DS $k_\gamma$. We see that the larger the DS becomes, the deeper the minimal value of $|s|$ is, which saturates for $k_\gamma\to\infty$
(see the main panel). The significant deviations of the unitarity are to be expected when $k_\gamma \gtrsim q_B$ and $\gamma_0 \gtrsim E_B$. The increase of the DS leads
to a fast increase of the threshold value of $\gamma_{0, \mathrm{min}}$. Comparing this to the Fig.~\ref{fig:S}a, we see that to reach the Zeno limit when $|s|\approx1$
requires an order of magnitude of increase of $\gamma_0$ as compared to $\gamma_{0,\mathrm{min}}$. 
We note that in this figure, $k_F =0 $ and the value of $\omega$ is set to its limiting case $\omega_\mathrm{th}$. For larger $\omega$ the
value of $|s|$ increases, and similarly $|s|$ grows with~$k_F$.}

%
%

\begin{figure}[htp]
  \centering
  \includegraphics[width=0.9\columnwidth]{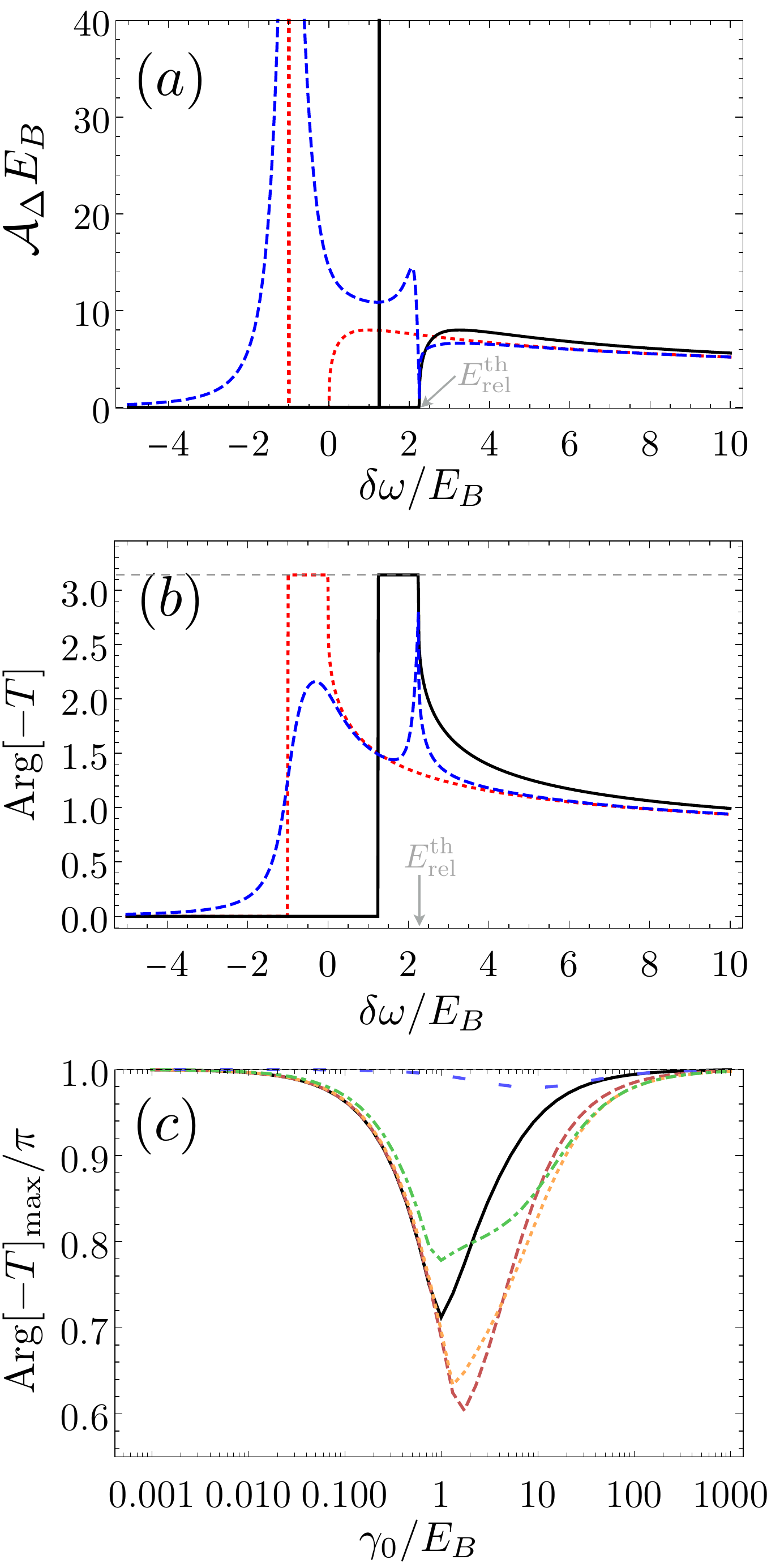}
  \caption{ 
    (a) Spectral function of the molecule $\mathcal{A}_\Delta = -2\mathrm{Im}[T(\bk,\omega)]$ calculated for $k=0$ and $\omega = k^2/2M + \delta\omega$; 
    here $k_F=0$, $k_\gamma=1.5 q_B$.
    The dotted red line -- the vacuum, lossless case, cf. Eq~\eqref{T0}.
    The dashed blue line -- $\mathcal{A}$ in the presence of moderate loss ($\gamma_0 = E_B$).
    The black line -- infinite loss case, cf. Eq.~\eqref{SED-zeno2}.
    The arrow indicates the threshold $E_\mathrm{rel}^{\mathrm{th}} = 2.25 E_B$.
    (b) The phase of the $T$-matrix, i.e., $\mathrm{Arg}[-T]$. The color-code and parameters are as in the panel (a).
    (c) The maximum value of $\mathrm{Arg}[-T]$, calculated with respect to parameter $\delta\omega$, 
    as a function of $\gamma_0$. The color-coding is the same as in Fig.~\ref{fig:S}a.
    In all the panels, $m_c=m_f$.}
\label{fig:argT}
\end{figure}

A further characterization of the scattering can be obtained by analyzing the phase of the $T$-matrix.  We compute $\mathrm{Arg}[-T(\bk,\omega)]$ for fixed $\bk$ as a
function of $\delta\omega$, where we define $\omega = k^2/2M + \delta\omega$. In Fig.~\ref{fig:argT} we consider the vacuum case at $k=0$. In absence of losses
(dotted-red lines in Fig.~\ref{fig:argT}a,b), the phase experiences a $\pi$-jump when crossing the bound state. Here, the phase is zero for $\delta\omega < -E_B$, and is
$\pi$ for $-E_B < \delta\omega <0$ , and for $\delta\omega>0$ it decays monotonically.

In presence of losses, as discussed above, the scattering threshold is shifted by $k_\gamma^2$, since impurities can escape only for momenta outside the DS. Therefore,
the phase of the $T$-matrix is analytical and the scattering phase shift defined only for $\omega>k_\gamma^2$ (see blue-dashed and solid black lines in
Fig.~\ref{fig:argT}a,b).

For finite loss rates (blue-dashed lines), the jump in the $T$-matrix-phase has a local maximum at the bound state but does not reach the maximal value $\pi$. By further
increasing the energy the $T$-matrix-phase then grows again to reach $\pi$ at the threshold value and then decreases monotonically and faster than in the lossless case
above threshold.

The $T$-matrix-phase behaves nontrivially upon increasing the loss rate. The double-peak structure present below threshold develops, in the quantum-Zeno limit of infinite
loss rate, into a single non-analytical plateau identical to the one present in the lossless case, apart from a blue shift of the threshold and the bound state.  As we
argued above, the vacuum scattering in the quantum-Zeno limit is the same as the lossless scattering in presence of a Fermi surface, where the role of $k_F$ is played by
$k_\gamma$. Consistently with this, the phase of the $T$-matrix in the quantum-Zeno limit and in particular the scattering phase shift indeed behaves as its lossless
counterpart in the presence of a medium, which has been studied in~\cite{nozieres1985bose,engelbrecht1990new,engelbrecht1992low}.

%
%

In particular, one can quantify the distance from the quantum-Zeno limit by the size of the jump of the $T$-matrix-phase across the bound state. This is shown in
Fig.~\eqref{fig:argT}c, where we plot $\mathrm{Arg}[-T(\bk,\omega)]$ maximized over $\delta\omega$. We see that for increasing $\gamma_0$, the maximum jump of the phase
of $T$ drops, and for $\gamma_0 \sim E_B$ it reaches a minimum, after which the further increase of $\gamma_0$ leads to recovery of the maximal jump of the phase.

%
%
\begin{figure*}[tb]
  \centering
  \includegraphics[width=\textwidth]{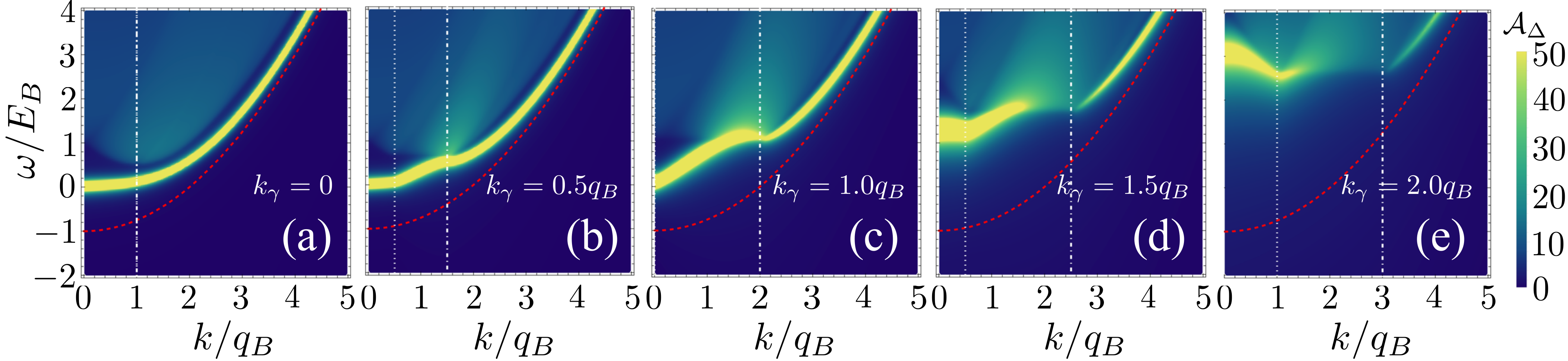}
  \caption{Molecular spectral function $\mathcal{A}_\Delta(\bk,\omega)$ (in units of $q_B^2 E_B$) for $m_c=m_f$ as a function of $|\bk|$ (in units of $q_B$) and $\omega$ (in units of $E_B$) for
    different sizes of the dissipative subspace: $k_\gamma/q_B = 0, 0.5, 1.0, 1.5, 2.0$ [from (a) to (e)]. The Fermi energy $\epsilon_F = k_F^2/2m_f$ is set to $k_F =
    q_B$, and $\gamma_0 = 20\, E_B$.  The dotted/dotted-dashed vertical lines indicate $|k_\gamma \mp k_F|$. The dashed red curve is the molecular free dispersion given by 
    $\omega = -E_B + k^2/2M$.}
  \label{fig:ADs}
\end{figure*}

The recovery of unitary scattering dynamics explains the re-appearance of an attractive-polaron branch for large loss rates shown in Fig.~\ref{fig:Ax}. As already
discussed in section \ref{sec-imp-SF}, this quantum-Zeno polaron features a clearly different dispersion affected by the presence of the additional scale $k_\gamma$, as
demonstrated by Fig.~\ref{fig:Axs}.

We now turn to the discussion of the interplay between $k_\gamma$ and the Fermi scale $k_F$, which is best understood by analyzing the molecular spectral function.  In
Fig.~\ref{fig:ADs} we show the spectral function $\mathcal{A}_\Delta$ for various values of $k_\gamma$ at a fixed density of fermions $k_F = q_B$. Moreover, we choose a
finite but large loss rate $\gamma_0 = 20 E_B$, where the scattering is already close to the quantum-Zeno regime.

For $k_\gamma=0$, i.e. without losses (see. Fig.~\ref{fig:ADs}a), the spectrum consists of a molecular bound state (in this dissipationless case, we artificially set a
finite lifetime for the bound state so that it is visible in the plot) and a continuum.  
\revM{In the non-self-consistent $T$-matrix approximation employed here, the boundaries of the continuum can simply 
be understood using energy and momentum conservation in the processes of Fig.~\ref{fig:FSDS}.}
This has been discussed already in previous
works \cite{engelbrecht1992low,schmidt2012fermi}, but we reproduce the arguments here since we will extend them below to the quantum-Zeno limit. A molecule with energy
$\omega$ and momentum $|\bk| = k$ can decay into the impurity and a fermion. The conservation laws require $\omega = \varepsilon_f(\bk') + \varepsilon_c(\bk'')$ and $\bk
= \bk'+\bk''$. The Pauli blocking sets an additional constraint $|\bk'|>k_F$.  For example, setting $k=0$, we have $\bk'=-\bk''$ and the energy conservation leads to
$\omega = |\bk''|^2/2m_r \leqslant k_F^2 / 2m_r$. Therefore, for $\omega \geqslant k_F^2 / 2m_r$ the molecule can decay into an impurity-fermion pair. Moreover, one can
show that for $k \approx k_F$, the threshold for the molecular decay behaves as: $k_F^2/2m_f + (k-k_F)^2/2m_c$. In general, for $k< m_c k_F/m_r$, the continuum boundary
has the form: $(k_F - m_r k/ m_c)^2/2m_r + k^2/2M$, and for $k> m_c k_F/m_r$ it is: $k^2/2M$, where $M = m_c+m_f$ is the total mass.  Consequently, for $k_\gamma = 0$,
the continuum has a minimum at $k=k_F$ and $\omega=\epsilon_F$. For smaller $k<k_F$, it is blueshifted due to Pauli blocking, and for $k=0$ it starts at $\omega =
k_F^2/2m_r$.

For a finite size of the DS and in the quantum-Zeno limit, several new features emerge in the molecular spectrum with respect to the lossless case. For small $k_\gamma$,
the spectrum is affected mostly in the vicinity of the Fermi momentum, namely in the region $k_F-k_\gamma < k < k_F + k_\gamma$, as can be seen from panels (b) and (c) in
Fig.~\ref{fig:ADs}. By increasing $k_\gamma$, the spectral weight of the continuum transfers to $\omega \geqslant k_\gamma^2/2m_c + k_F^2/2m_f$ in the region $k\approx
k_F$. Due to the presence of the DS, for smaller $k$ than $|k_F-k_\gamma|$ the spectrum is blueshifted more rapidly, and at $k=0$ it reaches $Q^2/2m_r$, where
$Q=\max(k_F,k_\gamma)$. Notice that, for $k$ lying in the interval $k_F-k_\gamma < k < k_F + k_\gamma$, the bound state dispersion is linear in $k$.  For even larger
$k_\gamma$, the effect of the Pauli blocking is enhanced by the presence of the DS, and the bound state energy is blueshifted even more. With increasing $k$ the bound
state energy drops, and enters the continuum for larger $k$.  Due to the forbidden regions in the formation of a molecule, the impurity and the fermion must have momenta
not smaller than $k_\gamma$ and $k_F$, respectively.  Therefore, the molecule with the smallest energy and momentum will have $k=|k_F-k_\gamma|$. Since the continuum
starts now at $k_\gamma^2/2m_c + k_F^2/2m_f$, the molecular state can appear at this energy also for $k=k_\gamma+k_F$.  This gives rise to the appearance of two minima in
the bound-state dispersion visible in Fig.~\ref{fig:ADs}e. This analysis makes apparent the crucial role played by the size of the DS in the quantum-Zeno regime, where
the molecule, and in turn, the polaron dispersion can be modified qualitatively.

\revM{The results presented in this section combined with Figs.~\ref{fig:Ax} and \ref{fig:Axs} highlight an important ingredient for the appearance of strong qualitative
effects of dissipation on molecular and polaron states. It is namely the molecular bound state which is strongly modified by the losses, while the continuum experiences
weaker modifications. This is reflected by the behavior of the polarons: only the attractive polaron is strongly affected by loss. This leads us to the expectation that
the phenomenology we propose here will crucially depend on the presence/absence of the molecular bound state and therefore on the dimensionality.}

\revM{Finally, we end this discussion by noting that all the results are obtained within the non-self-consistent T-matrix approach, whose limitations are known~\cite{schmidt2012fermi}. 
The full solution would require to employ self-consistency of the Green’s functions and would lead to modification of the molecular spectral function shifting
the position of the molecular and polaron resonances and affecting their lifetimes. These corrections, however, in the configurations and regimes considered affect the
quantities of present interest only quantitatively and do not change the overall physical picture.}

\section{Implementation of the model}
\label{sec:realization}

%
%
\begin{figure*}[tb]
\centering \includegraphics[width=\textwidth]{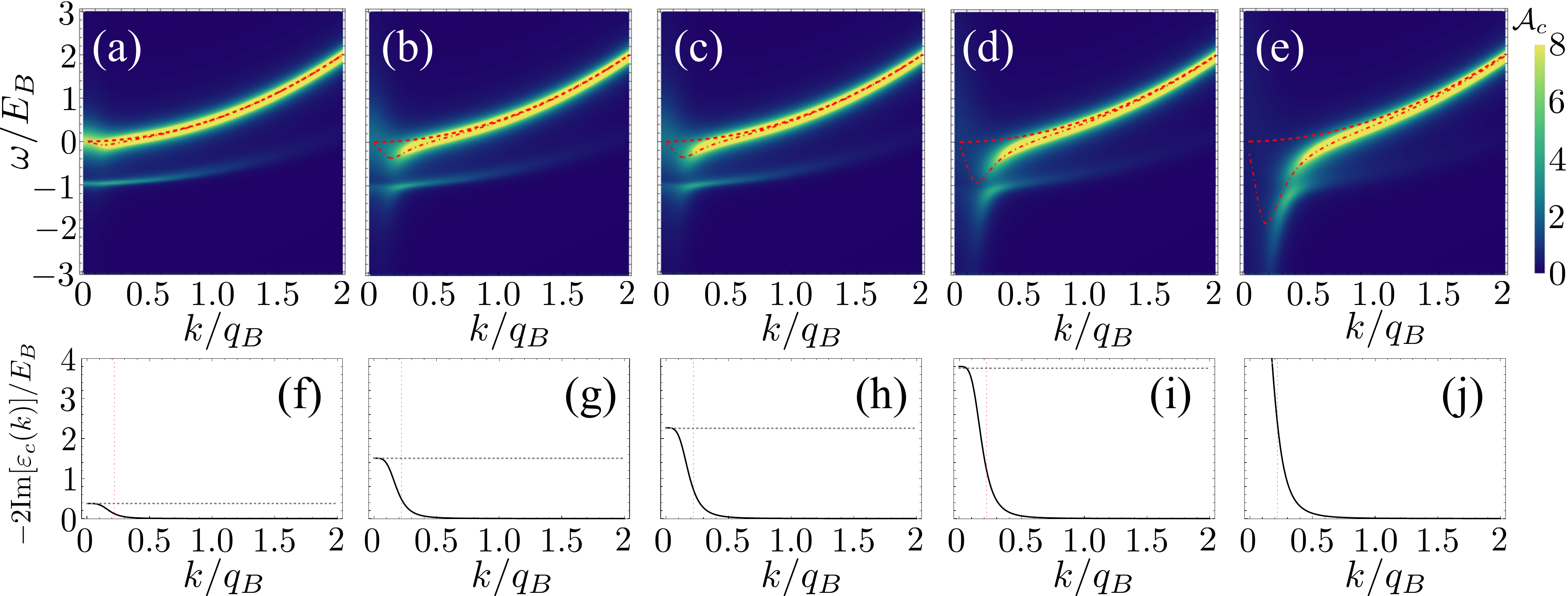}
  \caption{(Upper row) The spectral function $\mathcal{A}_c(\bk,\omega)$ of the impurity polariton and (lower row) its decay rate (without interactions with the fermions)
    for different Rabi frequencies $\Omega_R$: $\Omega_R/E_B = 5.59, 11.18, 13.69, 17.68, 25.00$ [from (a) to (e) and from (f) to (j)]. The Fermi energy $\epsilon_F =
    k_F^2/2m_f$ is set by $k_F = 0.22 q_B$. The polariton mixes an excitonic state $c$ (or stable internal state in the atomic context) with a photonic mode $d$ (or
    short-lived internal state in the atomic context). The short-lived state is characterized by the decay rate $\gamma_d = 333\, E_B$ and mass $m_d/m_c = 10^{-4}$; the
    fermion mass is $m_f=m_c$. The offset is set to zero, $\delta = 0$.  In the panels (a)--(e), the dashed red line is the dispersion relation $\epsilon_c(k) =
    k^2/2m_c$ of the exciton, the dotted-dashed red line shows the lower polariton branch $\mathrm{Re}[\varepsilon_c(k)]$, which for large $k$ approaches $\epsilon_c(k)$.
    In the panels (f)--(j), the dotted vertical red line is $k_F$, and the dotted horizontal gray line is $4\Omega_R^2/\gamma_d$.
}\label{fig:pos_m}
\end{figure*}

In this section, we discuss how the physics discussed so far could be realized in experiment. Starting from a setup featuring a density-density interaction between
fermions and impurity particles, the goal is to realize momentum-dependent losses for the impurities. The loss profile needs to single out a given region of momentum
space by being much larger there than elsewhere.

Recently it was demonstrated that it is possible to explore the two-dimensional Fermi polaron problem in transition metal dichalcogenide monolayer semiconductor embedded
into microcavities~\cite{sidler2017fermi,tan2019optical}.  In this type of setup, excitons are deeply bound and hybridize with the cavity photons, forming polaritons. By
optically exciting a small number of polaritons it is possible to create impurities that feature a mass that is much smaller than that of the surrounding electrons, since
photons propagate much faster. More important for the present discussion is the fact that the photons feature roughly momentum-independent losses due to mirror
leakage. These are inherited by the polaritons in a momentum-dependent fashion, as the photonic admixture indeed depends on momentum. This means that for one of the two
polaritons the loss rate will decrease by increasing momentum, while the opposite is true for the other polariton.

The simplest model for the above scenario is given by the following Hamiltonian
\begin{eqnarray}
  H_\mathrm{eff} &=& \sum_\bk \bigg\{\epsilon_c(\bk) \hat c^\dagger_\bk \hat c_\bk + [-\delta + \epsilon_{d}(\bk)] \hat d^\dagger_\bk \hat d_\bk + \\
  &&  + \Omega_R (\hat c^\dagger_\bk \hat d_\bk + \hat d^\dagger_\bk \hat c_\bk  )\bigg\},
\end{eqnarray}
where $\epsilon_{c,d}(\bk)$ is the dispersion of the exciton and the photon, respectively, and $\delta$ the relative energy offset.  $\Omega_R$ is the Rabi frequency of
interconversion between exciton and photon. The latter is subject to losses at a rate $\gamma_d$, corresponding to the Lindblad operator
\begin{equation}
  \mathcal{L}_\mathrm{diss} = \sum_\bk \gamma_d D[\hat d_\bk].
\end{equation}

The model just introduced also describes atoms featuring a (meta)stable internal electronic state coupled by a laser to a short-lived excited state, so that $\delta$ is the detuning between the laser frequency and the atomic transition frequency. A situation where the short-lived state has also a much smaller mass, as is naturally the case for excitons and photons, can be implemented with atoms by trapping the stable state in an optical lattice while leaving the excited state untrapped.

The Lindblad master equation resulting from $H_\mathrm{eff}$ and $\mathcal{L}_\mathrm{diss}$ translates in the Keldysh approach to the inverse retarded GF:
\begin{equation}
  \revM{[G^R_{cd}]^{-1}\!\! =}\!\! \left( 
  \begin{array}{cc}
    \omega\!-\! \epsilon_c(\bk) \!+\!i0 & -\Omega_R \\
    -\Omega_R & \omega \!-\! [\epsilon_d(\bk) \!-\!\delta ] \!+\! i \gamma_d/2 \!+\!i0
  \end{array} \right).
\end{equation}
Diagonalization of this matrix leads to the dispersion relation and loss rate of the polaritons. One of the complex eigenvalues, which we denote by $\varepsilon_c(\bk)$,
converges to $\epsilon_c(\bk)$ for large $|\bk|$; the other converges to $\epsilon_d(\bk)+i\gamma_d/2$ for large $|\bk|$ and we denote it with $\varepsilon_d(\bk)$. The
real part of the eigenvalue $\varepsilon(\bk)$ describes the position of the resonance, whereas its imaginary part characterizes its width, and hence the loss rate. For
simplicity, we assume that the low momenta are sufficient to describe the dynamics so that we may expand $\epsilon_{c,d}(\bk) \approx k^2/2m_{c,d}$, where $m_c$ and $m_d$
are the effective masses of the uncoupled particles. Now, by changing the ratio $m_c/m_d$, the detuning $\delta$ and Rabi frequency $\Omega_R$, one can adjust the regions
in momentum space where the lossy state $d$ interacts with the long-lived state $c$, thereby tuning the loss-profile of the polaritons.

In the actual numerical calculations, discussed below, we employ the same set of Dyson's equations introduced in section \ref{noneqGF} where the impurity dispersion and
loss rate are the one of the $c$-polariton (the one showing a loss rate which is larger at small momenta), which corresponds to replacing $\epsilon_c(\bk)+i\gamma(\bk)/2$
with $\varepsilon_c(\bk)$ in $G_0^R$ of equation \eqref{gc}. Here we are assuming that only one of the two polariton branches is excited and neglecting the mixing between
the branches induced by the interaction with the electrons. We are also neglecting the fact that the polariton-electron coupling strength depends on momentum. The
inclusion of these features, which in general leads to appreciable quantitative differences, is however beyond the scope of the present work.

In Fig.~\ref{fig:pos_m}, we show the impurity spectral function $\mathcal{A}_c(\bk,\omega)$ for different Rabi frequencies [see the panels (a)--(e)] $\Omega_R/E_B = 5.59,
11.18, 13.69, 17.68, 25.00$.  The photon (or short-lived state in the atomic context) has a large decay rate $\gamma_d = 333\, E_B$ and a small mass $m_d/m_c = 10^{-4}$;
the detuning is set to $\delta = 0$. In this parameter regime, the maximum decay rate is approximately given by $2\Omega_R^2/\gamma_d$.  The parameters are chosen in such
a way to have a steep dispersion relation $\epsilon_d(k)$ compared to $\epsilon_c(k)$, so that the tails of the polariton loss-profile drops on a scale smaller than
$E_B$. This feature is important to have a well defined DS and thus an almost forbidden region at large loss rates. The impurity and bath fermions are of the same
mass. The Fermi wavevector is $k_F = 0.22 q_B$. In the panels (f)--(j), we show the polariton loss-profiles corresponding to the panels (a)--(e). At small Rabi frequency,
the polariton dispersion shows minimal deviations from the original exciton dispersion and has also negligible losses, so that the two polaron-polariton peaks are
essentially the same as in Fig.~\ref{fig:Ax}a. By increasing the Rabi frequency the impurity-polariton dispersion develops a pronounced minimum due to the photon
admixture. The latter at the same time induces polariton losses concentrated around zero momentum. At sufficiently large $\Omega_R$ the quantum Zeno effect creates a
forbidden region and the dispersion of the polarons is significantly altered.  Upon entering the quantum-Zeno regime, both polaron-branches show an inversion of their
curvature i.e. of the polaron effective mass from positive to negative. In particular, for the attractive-polaron the negative curvature becomes very large in the closest
vicinity of the forbidden region, which seems to result from a mutual repulsion between the latter and the polaron branch.

Since the quantum-Zeno Fermi-polaron branch has weight only for finite momenta lying outside the forbidden region, the excitation of finite-momentum states is necessary
for experimental probing. In transition metal dichalcogenide monolayer semiconductors, optical excitation cannot transfer momenta which are comparable with the Fermi
momentum \cite{sidler2017fermi}, so that a direct probe of the attractive-polaron branch at large-enough momenta seems unfeasible. This portion of the attractive-polaron
branch is however accessible indirectly via the intermediate excitation of the repulsive-polaron branch at zero momentum and subsequent electron-mediated decay into the
attractive branch at momenta larger than $k_F$.  For ultracold atoms, the radiofrequency-spectroscopy
\cite{schirotzek2009observation,kohstall2012metastability,koschorreck2012attractive,thomas_polaron_2012} usually employed for probing Fermi-polarons cannot be used here
if one wants to excite finite-momentum states. Instead, two-photon Raman transitions \cite{shkedrov2019situ} can be exploited for a direct excitation of the
attractive-polaron branch at a given finite momentum.


\section{Conclusions}
\label{sec-conc}

In this work, we have studied the 2D Fermi-polaron problem in presence of impurity loss and drive. Our approach, employing diagrammatic methods for non-equilibrium Green's
function on the Keldysh time-contour~\cite{kamenev2011field,sieberer2016keldysh}, 
\revM{allows to consistently take into account dissipation and pump and study the strongly dissipative limit}.
\revM{It provides a framework to determine non-equilibrium distributions of fermions, molecules and impurities. 
We emphasize they cannot be determined on phenomenological grounds basing on equilibrium quantum field theory and require self-consistent treatment 
involving the Keldysh component of the Green functions. We leave the investigation of these intrinsically non-equilibrium features to future work.}

\revM{In the strongly dissipative regime,} we observed the reemergence of unitary Fermi-polaron-dynamics, but with an additional length scale set by the size of the region
forbidden by the quantum-Zeno effect. The nontrivial interplay between this scale and the Fermi scale induces large qualitative modifications of the impurity-fermion
scattering and consequently of the polaron dispersion.
We expect this phenomenology to be generic for situations where the losses single-out a region of momentum space. In particular, we have shown this to be naturally the case for exciton-polariton impurities and also possible to engineer for atomic gases.

In the same way as the polaron concept has been instrumental in developing a systematic understanding of fundamental phenomena in quantum many-body physics, polarons
undergoing a Quantum-Zeno projection provide a step towards the understanding of the many-body physics in presence of strong measurements. In particular, our findings
identify the emergence of a new scale associated with the Zeno-forbidden region in an interacting system as a fundamental and generic ingredient able to give rise to
novel collective phenomena.



\appendix

\revM{
\section{Non-equilibrium Green's functions}
\label{app-GFs}

\begin{table}
  \caption{\label{table}Diagrammatic representation of the Green's functions for the impurity (left column), fermions (middle column) and molecules (right column). To
    each line energy and momentum variables are attached, which are taken as the argument of the corresponding Green's function. The double arrows on the
    molecular lines indicate their composite nature. Keldysh GFs are drawn with solid lines, whereas solid-dashed (dashed-solid) lines represent
    retarded (advanced) Green functions.
  }
    \begin{tabular}{c c c}
      \hline\hline
      Impurity & Fermion & Molecule  \\  \hline
      \begin{minipage}[c]{0.3\columnwidth}\includegraphics[width=0.9\columnwidth]{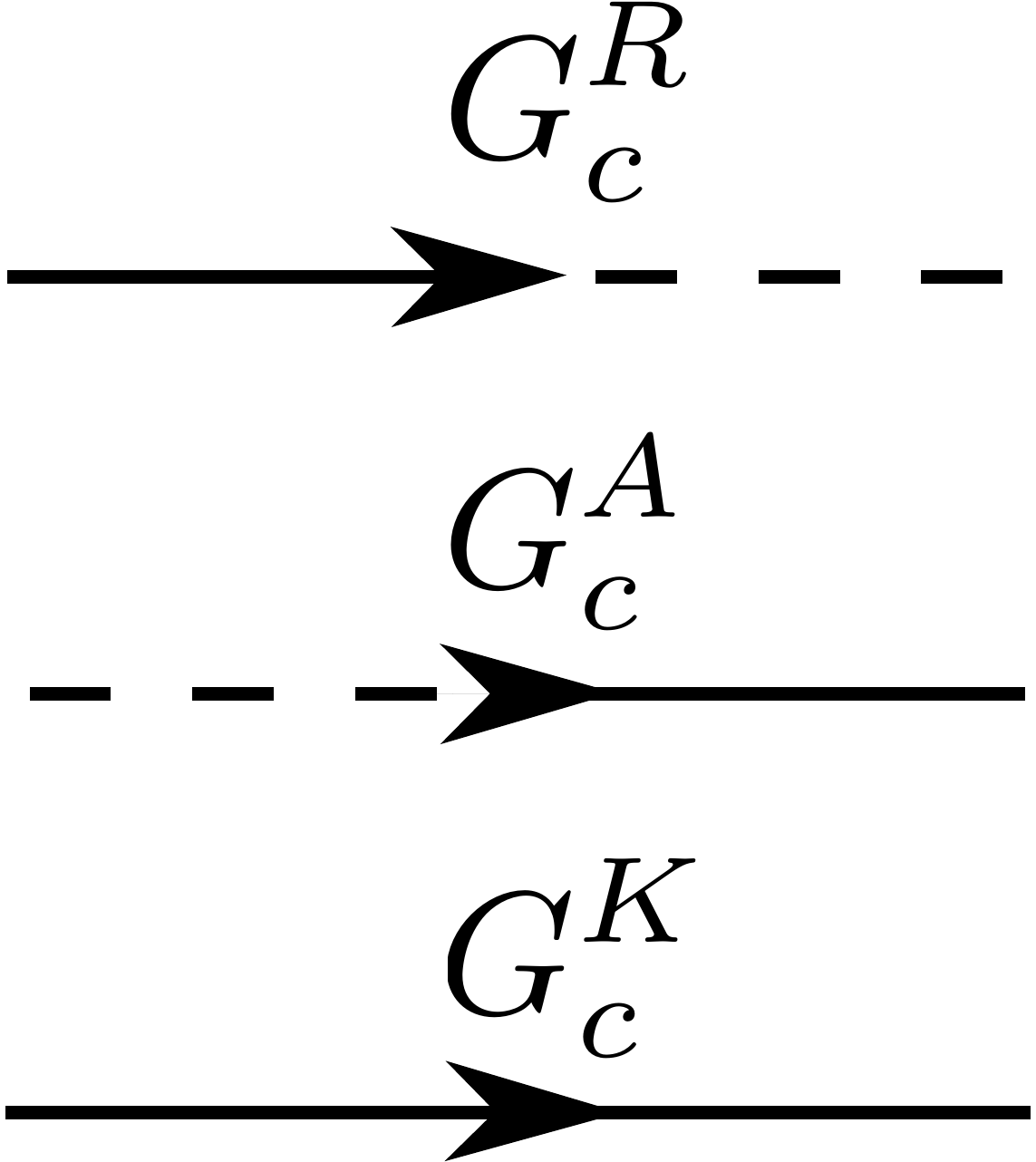}\end{minipage}  & 
      \begin{minipage}[c]{0.3\columnwidth}\includegraphics[width=0.9\columnwidth]{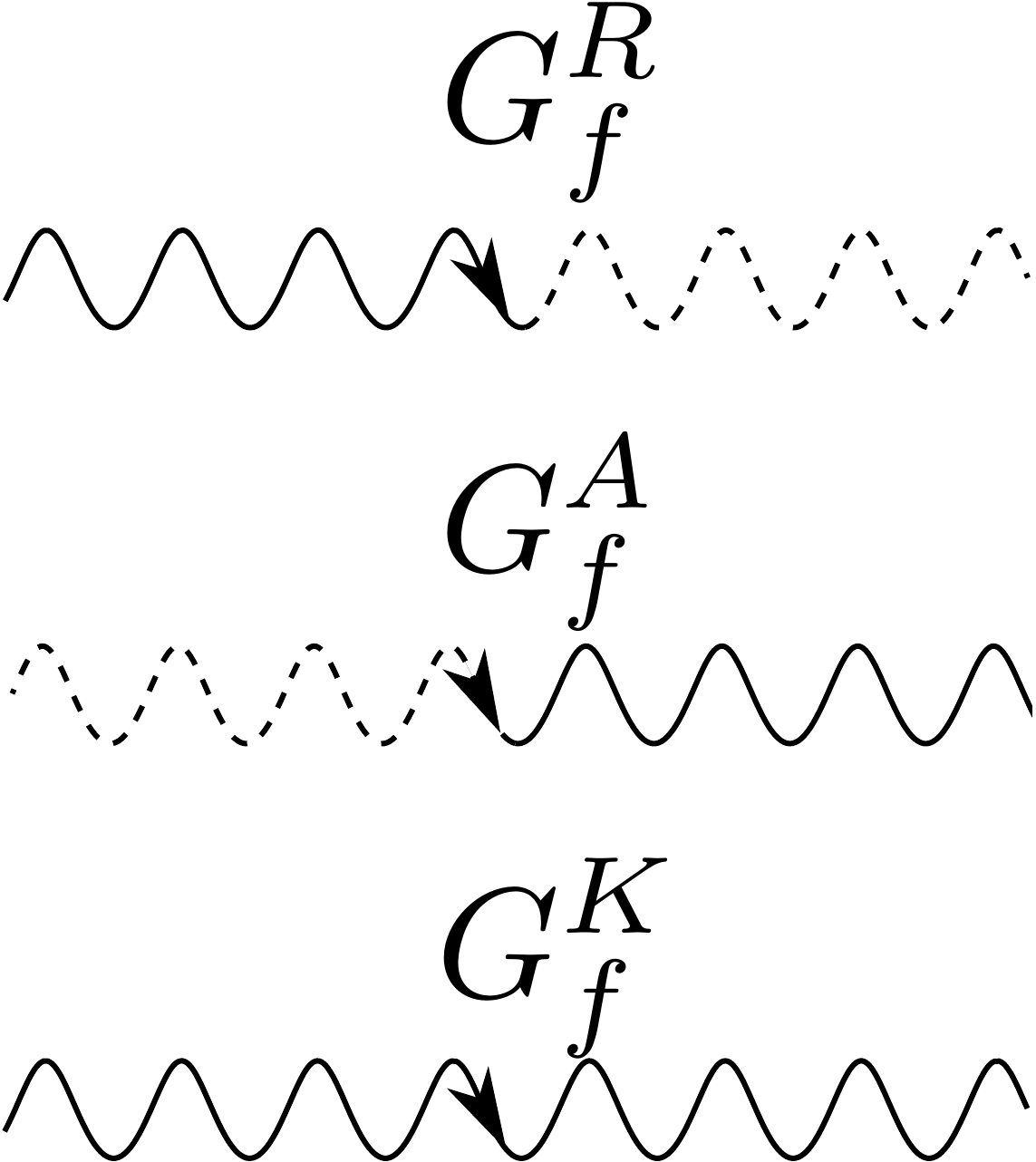}\end{minipage}  & 
      \begin{minipage}[c]{0.3\columnwidth}\includegraphics[width=0.9\columnwidth]{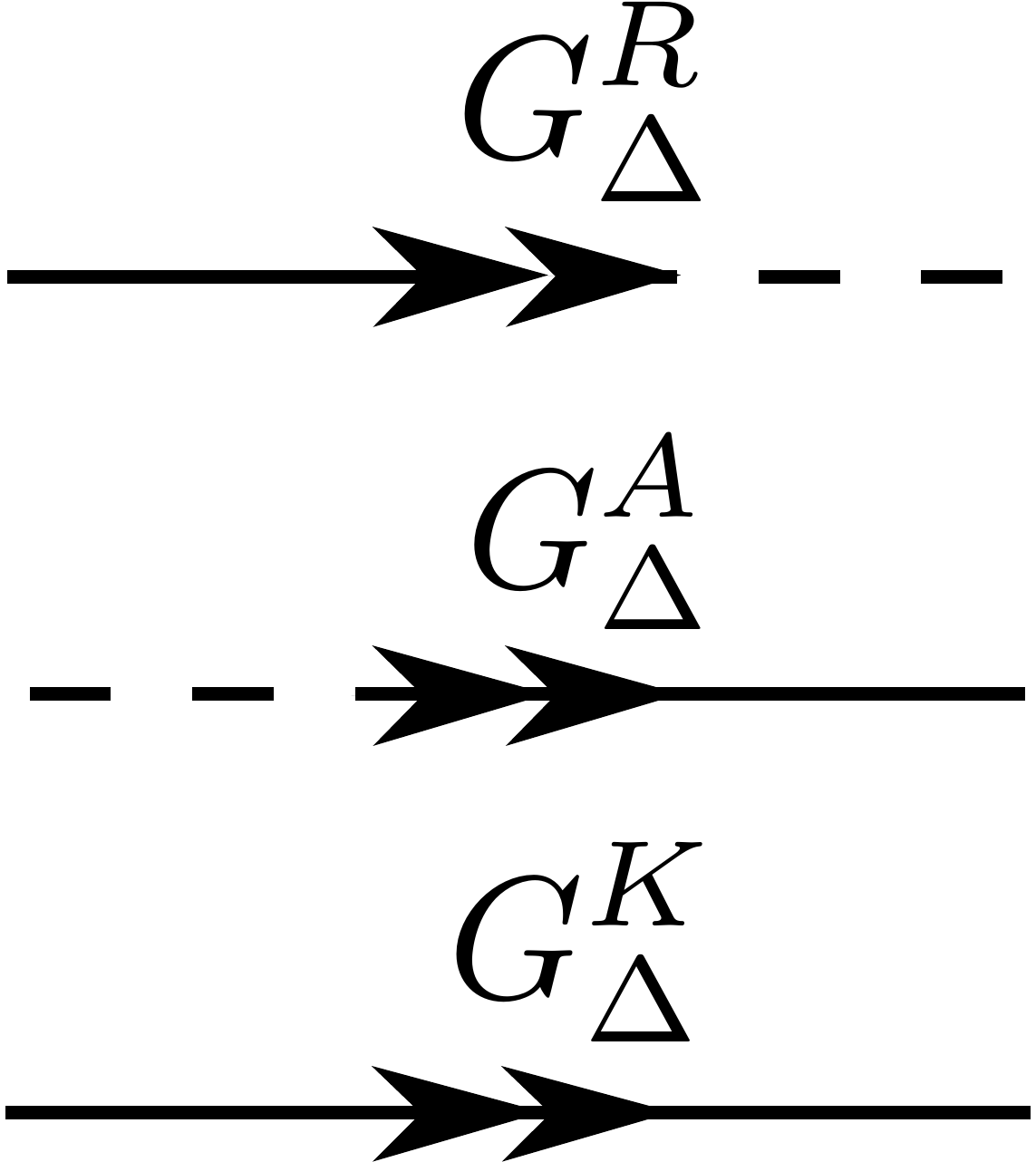}\end{minipage}  
      \\
      \hline\hline
    \end{tabular}
\end{table}

\subsection{Impurity, fermions and pairing}

\revM{The non-equilibrium Green's functions (GFs) are the building blocks of our theory. For the details of the theory in the path-integral formulation, and some calculations,
we refer the reader to Appendix~\ref{app-keldysh}.}  The GFs are defined by the relations~\cite{kamenev2011field}:
\begin{subequations}
  \label{defGFs}
  \begin{eqnarray}
    \label{GR}
    iG^R_\alpha(x,x') &\equiv& \theta(t-t') \av{ \big[ \hat \alpha(x), \hat \alpha^\dagger(x') \big]_\mp }, \\
    \label{GK}
    iG^K_\alpha(x,x') &\equiv&  \av{ \big[ \hat \alpha(x), \hat \alpha^\dagger(x') \big]_\pm }, 
  \end{eqnarray}
\end{subequations}
where the subscript $+/-$ indicates the anti-commutator/commutator, and $\hat{\alpha}^\dag$ creates a particle in one of the degrees of freedom, that is, for a
fermion $\hat{\alpha}^\dag=\hat{f}^\dag$ or for the impurity $\hat{\alpha}^\dag=\hat{c}^\dag$; hereafter, we denote the space-time variable by $x = (\x, t)$.  The
correlator in \eqref{GR} is the retarded GF, which is non-zero only for $t>t'$. Related to $G^R$ is the advanced GF denoted by
$G^A_\alpha(x,x')=[G^R_\alpha(x',x)]^*$. The correlator in \eqref{GK} is the so-called Keldysh GF or statistical propagator.  While the retarded GF $G^R$ contains
information only about the excitation spectrum of the system (being the expectation value of a (anti)commutator for bosons (fermions) it is indeed normalized to 1), the
Keldysh GF also depends on the occupation of excitation modes.  The Keldysh GF is in general independent of $G^{R/A}$ and has to satisfy its own evolution equation. Only
in thermal equilibrium the fluctuation-dissipation relation (FDR) allows to express the value of $G^K$ in terms of $G^R$ and $G^A$. It is only in this case that the
latter are sufficient for the description of the problem. Away from thermal equilibrium, the introduction of the Keldysh GF is necessary and in particular in
driven-dissipative systems even in stationary states~\cite{sieberer2016keldysh}.

In our system only the impurities are subject to loss and gain, while the fermions are stable particles. Thus, since the number of impurities is very small compared to
the density of fermions, we can assume that the fermions remain in a thermal equilibrium state. The Keldysh GF can then be related by FDR to the remaining two GFs as
follows:
\begin{equation}
  \label{gf-f}
  G^K_f(p) = F_f^\mathrm{eq}(\omega)[ G^R_f(p) - G^A_f(p)],
\end{equation}
with the thermal distribution $F_f^\mathrm{eq}(\omega) = 1 - 2 n_f(\omega)$ where $n_f(\omega)=\theta(\epsilon_F - \omega)$ is the Fermi-Dirac distribution at $T=0$,
$\theta$ the unit step function, and $\epsilon_F$ the Fermi energy. Using the time- and space-translation invariance in the steady state, all GFs are expressed by their
Fourier transforms in the relative coordinate, for example, $G^R_f(x,0) \equiv G^R_f(\x,t) \equiv G^R_f(x) = \sum_p e^{i p\cdot x}G^R_f(p)$, and hereafter, following
Ref.~\cite{kamenev2011field}, we denote the momentum-energy vector $p=(\k,\omega)$, and the scalar product $p\cdot x = \bk\cdot\x - \omega t$; the sum here is $\sum_p =
\sum_\bk \int\!\! d\omega/2\pi$.  Finally, the retarded (advanced) GF is given by $G_f^{R/A} = (\omega - \varepsilon_f(\bk) \pm i 0)^{-1}$, where the infinitesimal
imaginary number, denoted with $\pm i0$, ensures the proper support in real time.

%
%
\begin{figure}[tb]
  \centering 
  \includegraphics[width=1.0\columnwidth]{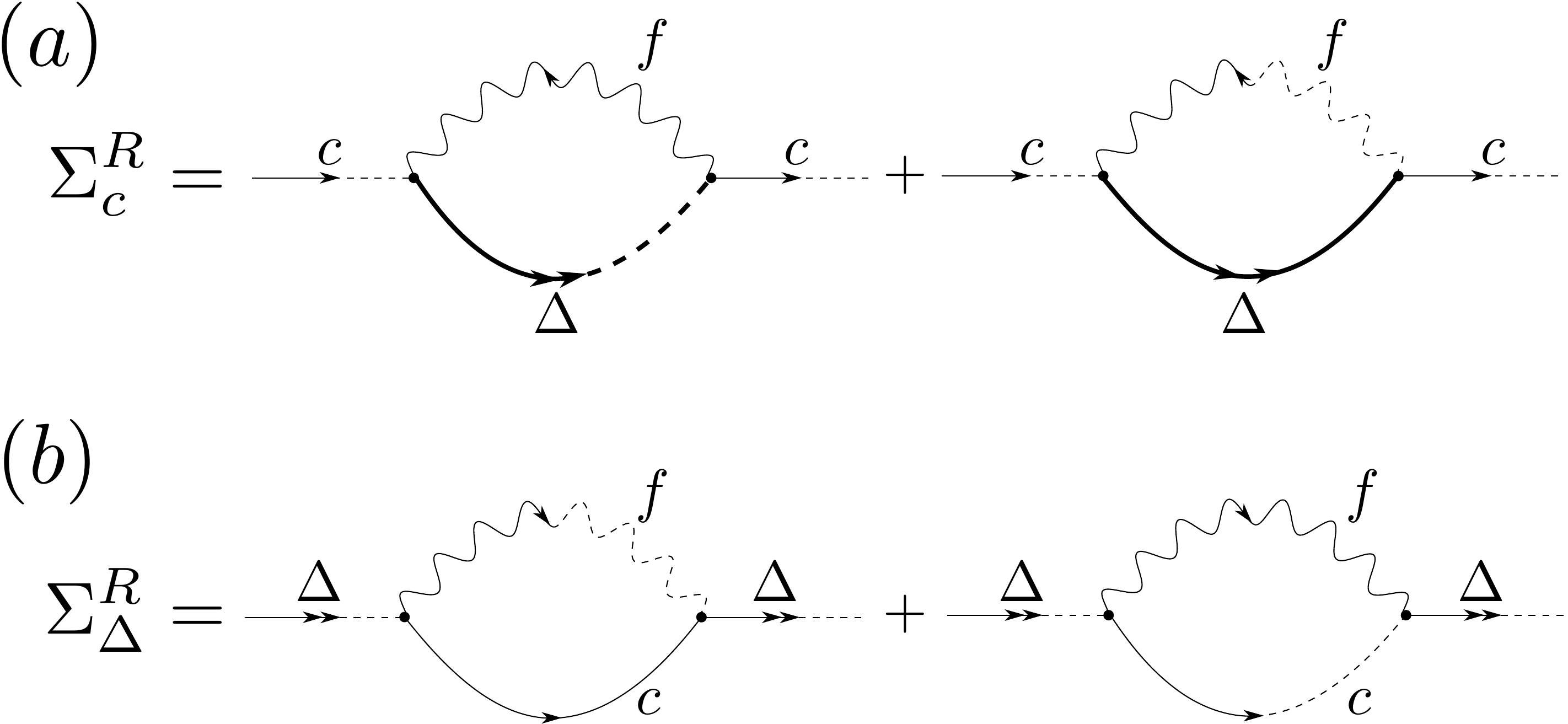}
  \caption{The diagrams showing the scattering processes that give rise to the impurity (a) and the molecule (b) self-energies (after truncation of external lines).  The
    Keldysh structure, i.e., the types of Green functions, is explicitly presented, cf. Tab.~\ref{table}.
    In (a), during propagation of the impurity, the particle scatters with a fermion $f$ to form a molecule $\Delta$, and then the molecule decays back into the fermion
    and the impurity. This process yields $\Sigma_c^R$.
    In (b), a propagating molecule decays into a fermion and the impurity, which subsequently collide to form the molecule. This process gives rise to
    $\Sigma_\Delta^R$. To highlight the compound molecular structure, we denote its propagator with the line with two arrows.}
\label{fig:diag_xD}
\end{figure}

Neglecting the impurity-fermion interaction, it is straightforward to evaluate the impurity GFs. These GFs, which are the building blocks for our diagrammatic approach,
are referred to as ``bare'' (denoted by an additional subscript $0$). Since we assume that the fermions are not renormalized by the presence of a small number of
impurities, we drop the subscript ``0'' from fermionic GFs.  The impurity GFs read
\begin{subequations}
  \label{gc}
  \begin{eqnarray}
    G_{0,c}^{R/A}(\bk,\omega) &=& \frac{1}{\omega - \varepsilon_c(\bk) \pm i \frac{\gamma(\bk)}{2}},\\
    G_{0,c}^K(\bk,\omega) &=& F_{0,c}(\bk) [G_{0,c}^R(p) - G_{0,c}^A(p)], 
  \end{eqnarray}
\end{subequations}
with $F_{0,c}(\bk) = 1 + 2 n_{0,c}(\bk)$ and $n_{0,c}(\bk) = \Omega(\bk)/\gamma(\bk)$. The distribution function $F_{0,c}$ has the same structure as $F_{f}^\mathrm{eq}$
(the opposite relative sign is due to the bosonic statistics for the impurities). However, in place of the thermal occupation the ratio of pump to loss rate
appears in $F_{0,c}$. This ratio gives the maximal number of impurities $n_{0,c}(\bk)$ allowed to occupy a given momentum state $\bk$.  

In the Fermi-polaron problem, bound-state formation between the impurity and a bath-fermion plays a prominent role~\cite{prokof2008fermi}. Thus, featuring a pole already in the two-body problem, the particle-particle scattering vertex dominates the interactions in the Fermi polaron problem. To capture this pairing channel we introduce an auxiliary molecular field, which
provides access to molecular degrees of freedom having the bound state as a long lived excitation. As already known from thermal equilibrium, this approach to the
interactions between impurity and bath is fully equivalent to the $T$-matrix approach.  As we show in Appendix~\ref{app-keldysh}, the molecular field can be properly
introduced within the Keldysh functional-integral formalism by performing a Hubbard-Stratonovich transformation.
This leads to the following interaction part of the functional-integral action:
\begin{align}
  S_{\Delta,f,c} = - \int_\mathcal{C} \! dt \int \! d\x \, ( \bar c \bar f \Delta + \bar \Delta f c),
\end{align}
where $\mathcal C$ is the Keldysh time-contour~\cite{kamenev2011field}.  The molecular field $\Delta$ introduced via the Hubbard-Stratonovich transformation has actually
no own dynamics, the latter being generated only by the interaction with impurities and fermions. Formally, this translates in the bare molecular propagator being simply
given by $G_{0,\Delta}^{R/A}=(1/U \pm i0)^{-1}$. The non-interacting Keldysh component of the molecular action is vanishingly small and is overwritten by the interaction
corrections as expressed by the self-energies, which we introduce next.

\subsection{Dyson equation and diagrammatic expansion}
\label{sec-dyson}

The interactions in the system lead to the modification (so-called dressing) of the bare GFs. We compute these corrections from Dyson's equation, for details see
Appendix~\ref{app-dyson}. Within the Keldysh formalism, one has to solve two coupled Dyson's equations for the retarded and the Keldysh GFs, respectively, since, as
explained above, the two are in general independent when the out-of-equilibrium case is considered.  In what follows, we are interested in the spectral signatures of the
polarons, which are contained in the retarded, dressed GF of the impurity
\begin{equation}
  G_c^R(\bk,\omega) = \frac{1}{\omega - \varepsilon_c(\bk) + i \frac{\gamma(\bk)}{2}  - \Sigma^R_c(p) + i0}.
\end{equation}
Here the retarded self-energy $\Sigma^R_c$ results from the interactions, the relevant process being the one where the impurity scatters with a fermion, temporarily forms
a molecule, which then decays back into an impurity and a fermion. This process is displayed in Fig.~\ref{fig:diag_xD}a in terms of Feynman diagrams, where, as
summarized in Tab~\ref{table}, each line corresponds to a GF and the intersection points to interaction vertices. In particular, dressed and bare GFs are depicted as
thick and thin lines, respectively. The choice of the diagrams in Fig.~\ref{fig:diag_xD} is a direct generalization of a non-self-consistent $T$-matrix approach to a
driven-dissipative setting. In particular, if we assume thermal equilibrium, i.e., set the impurity losses to zero, it reproduces the result of the non-self-consistent
$T$-matrix approach in the impurity limit.

Every diagram corresponds to a mathematical expression for the self-energy that can be obtained as follows. To each line an energy-momentum vector is assigned. For
internal lines, the latter is taken as the argument of the GF, while for the external lines it sets the argument of the self-energy we are computing. At each vertex
energy and momentum conservation is imposed.  Finally, a sum is performed over all energy-momentum vectors of the internal lines.

With this procedure, we obtain from Fig.~\ref{fig:diag_xD}a the following self-energy:
\begin{equation}
  \label{SEx}
  \Sigma_c^R(p) =\! -\frac{i}{2 V}\! \sum_{p'}\bigg[ G_f^K(p')G_\Delta^R(p+p') + G_\Delta^K(p') G_f^A(p'-p) \bigg],
\end{equation}
where $V$ is the area of the two-dimensional system. The advanced/retarded GF of each particle is always multiplied by a Keldysh GF of the other particle, which carries
information about occupations of the states.

In order to compute the impurity self-energy from Eq.~\eqref{SEx}, we need the molecular GF. The retarded component is given by
\begin{equation}
  \label{eq:molecule_gf-a}
  G_\Delta^R(p) = \frac{1}{- \nu^2 - \Sigma_\Delta^R(p)+i0},
\end{equation}
with $\nu^2 = -1/U$ and $\Sigma_\Delta^R(p)$ the molecular self-energy. The molecular Keldysh GF can be parametrized again as $G_\Delta^K = F_\Delta (G^R_\Delta -
G^A_\Delta)$. If the number of impurities is very small, the number of molecules is also very small, resulting in $F_\Delta\approx 1$.  In the impurity limit and assuming
the Fermi-bath to be at $T=0$, we thus have
\begin{equation}
  \label{SEx2}
  \Sigma_c^R(\bk,\omega) \!\!=\!\! \frac{1}{V}\!\! \sum_{\bk' \in \mathrm{FS}}\frac{1}{-\nu^2 \!-\! \Sigma_\Delta^R(\bk\!+\!\bk', \omega \!+\! \varepsilon_f(\bk')) + i0},
\end{equation}
where the summation is restricted to momenta from inside the Fermi Sea (FS) for which $|\bk'| \leqslant \epsilon_F$. This constraint on the fermion momenta is a direct consequence
of the presence of $G_f^K$ in Eq.~\eqref{SEx}. Physically, during the propagation of the impurity a transient molecular state can be formed if the impurity collides with 
a fermion, and this can be taken only from the Fermi Sea.

The molecular self-energy $\Sigma_\Delta^R(p)$ is needed to evaluate $\Sigma_c^R(p)$ and in turn $G_c^R(p)$. The dominant process that leads to the modification of the
propagation of a free molecule is depicted in Fig.~\ref{fig:diag_xD}b. A propagating molecule decays virtually into an impurity and a fermion, which subsequently collide
to form the molecule again. The expression for the self-energy resulting from the Feynman diagram reads
\begin{equation}
\label{SED}
  \Sigma_\Delta^R(p) \!=\! \frac{i}{2V}\!\sum_{p'}\!\bigg[ G_{0,c}^K(p')G_f^R(p-p') + G_f^K(p') G_{0,c}^R(p-p') \bigg],
\end{equation}
where we use the bare impurity GFs.  Since all functions entering the equation for $\Sigma_\Delta^R$ are known analytically, the self-energy can be computed explicitly in
the impurity limit:
\begin{equation}
\label{SED2-a}
  \Sigma_\Delta^R(\bk,\omega) \!\!=\!\! 
  \frac{1}{V} \!\! \sum_{\bk' \notin \mathrm{FS}} \frac{1}{\omega\! -\! \varepsilon_c(\bk - \bk') \!-\! \varepsilon_f(\bk') \!+\! i \frac{\gamma(\bk - \bk')}{2}\! +\! i0}.
\end{equation}
Notice that the sum over momenta extends outside the Fermi Sea. This is a consequence of the fact that the propagating molecule can decay into an impurity-fermion pair,
but due to Pauli blocking, the fermion cannot occupy any state from the Fermi Sea.  Due to the composite nature of the molecule, its GF is affected by impurity loss. In
turn, the impurity self-energy in Eq.~\eqref{SEx2} is affected by loss through $\Sigma_\Delta^R$.  Since $G_{0,c}^K(p')$ does not depend on $\Omega(\bk)$ in the impurity
limit, there is no pump-dependence left in any of the retarded GFs, {which are discussed in the main text.}

Finally, let us make more explicit the connection between the above results and the known case of the lossless Fermi polaron. For this purpose, let us consider
the quasi-particle excitations of the system, whose energy and damping is given by the complex poles of the retarded impurity GF. By setting $\gamma(\bk)$ to zero we
obtain the following equation determining the quasi-particle dispersion: $\omega(\bk) = \varepsilon_c(\bk) -\Sigma_c^R(\bk,\omega(\bk))$. Exactly the same
equation~\cite{chevy2006universal,combescot2007normal, massignan2014polarons,kroiss2014diagrammatic} is derived when the variational approach is employed with the Chevy
polaron ansatz $\ket{\Psi(\bk)} = \chi \hat c_\bk^\dagger \ket{\mathrm{FS}} + \sum_{\bk',\mathbf{q}} \chi_{\bk'\mathbf{q}}\hat c_{\bk-\bk'+\mathbf{q}}^\dagger \hat
f_{\bk'}^\dagger \hat f_{\mathbf{q}}\ket{\mathrm{FS}}$, where $\bk' \notin \mathrm{FS}$ and $\mathbf{q}\in \mathrm{FS}$. This trial wave function, with variational
parameters $\chi$ and $\chi_{\bk'\mathbf{q}}$, includes single particle-hole excitations and is able to capture the influence of the molecular state on the impurity
propagation.

}

\revM{
\section{Keldysh path-integral formulation of the theory}
\label{app-keldysh}
}

In our work, we follow the approach presented in Refs.~\cite{kamenev2011field} and~\cite{sieberer2016keldysh}.  The quantum average of any operator can be calculated
using the path-integral formulation.  Here, the quantum averages over a state, given by a density matrix, are calculated by summing over all possible realizations of
fields weighted by a complex exponent that can be interpreted as an action. Below, we present the structure of the theory and derive the formulas presented in the main
text.

We first consider the impurities. Although the statistics of the impurities is irrelevant, we choose them to be bosons, since this simplifies the functional-integral
formulation. We then treat the fermionic bath using Grassmann fields. Finally, we include the interactions between impurities and fermions, which can be reformulated by
introducing the molecular fields and their interaction vertex with fermions and impurities.

\subsection{The partition function}

Our starting point is the partition function of the system, written as an integral over field configurations:
\begin{equation}
  Z = \int \mathcal{D}[c,f] e^{i S[\bar c_{+}, c_{+}, \bar c_{-}, c_{-},  \bar f_{+}, f_{+}, \bar f_{-}, f_{-}] },
\end{equation}
where for simplicity in the integration measure $\mathcal{D}[c,f] \equiv \mathcal{D}[\bar c_{+}, c_{+}, \bar c_{-}, c_{-}, \bar f_{+}, f_{+}, \bar f_{-}, f_{-}]$ we
suppressed dependence on position (or momentum) and time of the fields. The action $S= S_c + S_f + S_{\mathrm{int}}$ can be decomposed into three parts that describe the
bosonic impurities, fermions and the interaction, respectively. Below, we describe them separately.

The bosonic impurities are governed by
\begin{eqnarray}
  S_c\!\! &=&\!\! \int\!\! dt\! \sum_\bk \bigg\{\bar c_+ [i \partial_t \!-\!\varepsilon_c(\bk)]c_+ \!-\! \bar c_- [i \partial_t \!-\!\varepsilon_c(\bk)]c_- \quad\\
  && - i [\gamma(\bk) + \Omega(\bk)] \bigg[ c_+ \bar c_- - \frac12( \bar c_+ c_+ + \bar c_- c_-)  \bigg]  \\
  && - i \Omega(\bk) \bigg[ \bar c_+  c_- - \frac12( \bar c_+ c_+ + \bar c_- c_-)  \bigg]\bigg\},
\end{eqnarray}
where the complex-valued field $c_\pm \equiv c_\pm(\bk,t)$ resides on the forward (backward) branch of the Keldysh contour~$\mathcal{C}$, which goes from $t= -\infty$ to
$t=\infty$ (forward branch) and then from $t=+\infty$ to $t=-\infty$ (backward branch). Next, we perform the Keldysh rotation according to
\begin{equation}
  c_{cl/q} = \frac{c_+ \pm c_-}{\sqrt{2}} \textrm{ and } \bar c_{cl/q} = \frac{\bar c_+ \pm \bar c_-}{\sqrt{2}},
\end{equation}
where the indices $cl$ and $q$ stand for ``classical'' and ``quantum'', respectively. Since we are interested in the stationary states, we perform the Fourier transform
and express the fields in terms of the frequency variable.  In the new basis, the action takes the form:
\begin{equation}
  S_c = \sum_p \bar c_\alpha (p) G_{0,c,\alpha\alpha'}^{-1}(p)c_{\alpha'}(p),
\end{equation}
where the sum over repeated indices is implied here and in the following. The object $G_{0, c,\alpha\alpha'}^{-1}(p)$ is the bare inverse of
the impurity GF, which reads:
\begin{equation}
\label{a-igc}
  G_{0, c}^{-1} = 
  \left( 
  \begin{array}{cc}
    0 & [G_{0,c}^A]^{-1} \\
    {[}G_{0,c}^R{]}^{-1} & {[}G_{0,c}^{-1}{]}^{K}
  \end{array} \right).
\end{equation}
We note that the $q-q$ term is the Keldysh component of the inverse of the GF, and its inverse is not equal to the Keldysh component of the GF (see below).
The components of the matrix are given by
\begin{subequations}
\begin{eqnarray}
{[}G_{0,c}^{R/A}{}]^{-1} &=& \omega - \varepsilon_c(\bk) \pm i\frac{\gamma(\bk)}{2}, \\
{[}G_{0,c}^{-1}{]}^{K} &=& i[\gamma(\bk) + \Omega(\bk)].
\end{eqnarray}
\end{subequations}
The inversion of Eq.~\eqref{a-igc} results in Eq.~\eqref{gc}. In this way, we have
\begin{equation}
\label{a-gc}
  G_{0, c} = 
  \left( 
  \begin{array}{cc}
    G_{0,c}^{K} & G_{0,c}^R \\
    G_{0,c}^A & 0 
  \end{array} \right),
\end{equation}
and the entries are given by
\begin{eqnarray}
  G_{0,c,\alpha\alpha'}(p) &=& \av{ c_\alpha(p) \bar c_{\alpha'}(p) }\\
  &=& \int\! \mathcal{D}[\bar c, c]\,  c_\alpha(p) \bar c_{\alpha'}(p) e^{i S_c[c, \bar c]}.
\end{eqnarray}
We emphasize that in the Keldysh formulation of the theory the $cl-cl$ component of $G_{0, c}^{-1}$ is zero. This property applies also to the full dressed GF when the
interactions are included and the average is taken over the total action~$S$.  This is referred to as the ``causal structure'' and originates from the detailed derivation
of the path-integral formalism~\cite{kamenev2011field}.  The causal structure of the theory is reflected in the vanishing of $q-q$ component of the GF.

Now, we proceed to the fermionic action given by
\begin{equation}
  S_c = \int_\mathcal{C}\!\! dt\, \sum_\bk \bar f(\bk,t) [i \partial_t - \varepsilon_f(\bk)] f(\bk,t).
\end{equation}
Here, the fields $f$ and $\bar f$ are independent Grassmann non-commuting numbers. Next, we write explicitly the fields on forward and backward branches of the Keldysh
contour, and perform the Keldysh rotation according to:
\begin{eqnarray}
  f_{\pm} = \frac{f_1 \pm f_2}{\sqrt{2}} \textrm{ and }   \bar f_{\pm} = \frac{f_2 \mp f_1}{\sqrt{2}}.
\end{eqnarray}
As a result, after taking the Fourier transform, we obtain the action:
\begin{equation}
  S_f = \sum_p \bar f_a(p) G_{f,aa'}^{-1}(p) f_{a'}(p),
\end{equation}
where the Keldysh indices $a$ and $a'$ can be either 1 or 2. The inverse of the fermionic GF is given by
\begin{equation}
\label{a-igf}
  G_{f}^{-1} = 
  \left( 
  \begin{array}{cc}
    {[}G_{f}^R{]}^{-1} & {[}G_{f}^{-1}{]}^{K} \\
    0 & [G_{f}^A]^{-1} 
  \end{array} \right),
\end{equation}
which, after inversion, leads to the following GF:
\begin{equation}
\label{a-gf}
  G_{f} = 
  \left( 
  \begin{array}{cc}
    G_{f}^R  & G_{f}^{K} \\
    0 &  G_{f}^A 
  \end{array} \right),
\end{equation}
and, assuming fermions of infinite lifetime, the entries are given by $G_f^{R/A}(p) = [\omega - \varepsilon_f(\bk) \pm i0]^{-1}$ and $G_f^K$ is in Eq.~\eqref{gf-f}. The
causal structure of the theory leads to $[G_f]_{21}\equiv0$.

\subsection{Interaction and molecules}

The action corresponding to the interaction in Eq.~\eqref{eq:Hint} is given by
\begin{equation}
\label{a-sint}
  S_\mathrm{int} = -U \int_\mathcal{C} \! dt \int \! d\x \, \bar c(x) c(x) \bar f(x) f(x).
\end{equation}
We proceed with the Hubbard-Strantonovich transformation that introduces the molecular field which mediates the contact interaction in Eq.~\eqref{a-sint}.

To this end, we introduce an auxiliary fermionic field $\Delta(x)$ that decouples the interaction potential:
\begin{equation}
e^{i S_\mathrm{int}} = \int\!\mathcal{D}[\bar\Delta,\Delta]e^{i S_\Delta + i S_{\Delta,f,c}  },
\end{equation}
with
\begin{subequations}
  \begin{eqnarray}
    S_\Delta &=& \int_\mathcal{C} \! dt \int \! d\x \, \bar \Delta(x) U^{-1} \Delta(x), \\
    S_{\Delta,f,c} &=& - \int_\mathcal{C} \! dt \int \! d\x \, ( \bar c \bar f \Delta + \bar \Delta f c).
  \end{eqnarray}
\end{subequations}
Now, the action is linear in the fermionic fields, and thus the fermionic degrees of freedom can be integrated out.
Consequently, we write
\begin{equation}
  \int\!\mathcal{D}[\bar f,f] e^{i S_f + i S_{\Delta,f,c}} = e^{i S_{\Delta c}},
\end{equation}
where the effective resulting action is given by
\begin{eqnarray}
  S_{\Delta c} &=& -\frac{1}{2} \int\! dx' \int\! dx' \bar c_\beta(x') \Delta_u(x) \gamma^\alpha_{a,u} \times \\
  && \times G_{f,ab}(x,x') \gamma^\beta_{b,v}\Delta_{v}(x')c_\alpha(x).
\end{eqnarray}
The sum over indices $a,b,u,v$ runs over $1$ and $2$, whereas the sum over $\alpha,\beta$ runs over $cl$ and $q$. The matrix $\gamma^{cl}$ is the identity matrix in
the Keldysh indices, i.e, $\gamma^{cl}_{a,b} = \delta_{a,b}$ and $\gamma^q$ has unit entries on the off-diagonal, i.e, $\gamma^q_{ab} = 1 - \delta_{a,b}$.
Finally, the molecular action $S_\Delta$ leads to the inverse bare molecular GF:
\begin{equation}
\label{a-iD}
  G_{0,\Delta}^{-1} = 
  \left( 
  \begin{array}{cc}
    [G_{0,\Delta}^R]^{-1}   &  [G_{0,\Delta}^{-1}]^K \\
    0 &  [G_{0,\Delta}^A]^{-1} 
  \end{array} \right),
\end{equation}
where $ [G_{0,\Delta}^{R/A}]^{-1} = 1/U \pm i0$.  The action in the Keldysh indices and in momentum space can be written as $S_\Delta = \sum_p \bar \Delta_a(p)
G_{0,\Delta,ab}^{-1}(p) \Delta_b(p)$; here $a,b$ takes values 1 or 2.  The off-diagonal Keldysh component is vanishingly small for the bare propagator, but it is
renormalized when the interactions are taken into account.  

Going back to the momentum representation the action reads
\begin{eqnarray}
  S_{\Delta c} &=& -\frac{1}{2 V} \sum_{p_1,p_2,p_3,p_4}\!\!\!\!\!\!\!{}'\ G_{f,ab}(p_1-p_4) \gamma^\alpha_{u,a} \gamma^\beta_{b,v} \times \nonumber\\
  && \times \bar \Delta_u(p_1)\Delta_v(p_2) \bar c_\beta(p_3) c_\alpha(p_4),
\label{a-SDc}
\end{eqnarray}
where the prime on the sum sign indicates the constraint $p_1+p_2 = p_3+p_4$; the sum over repeated discrete indices $a,b,u,v,\alpha,\beta$ is implicit.  

\begin{figure}[tb]
  \centering \includegraphics[width=1.0\columnwidth]{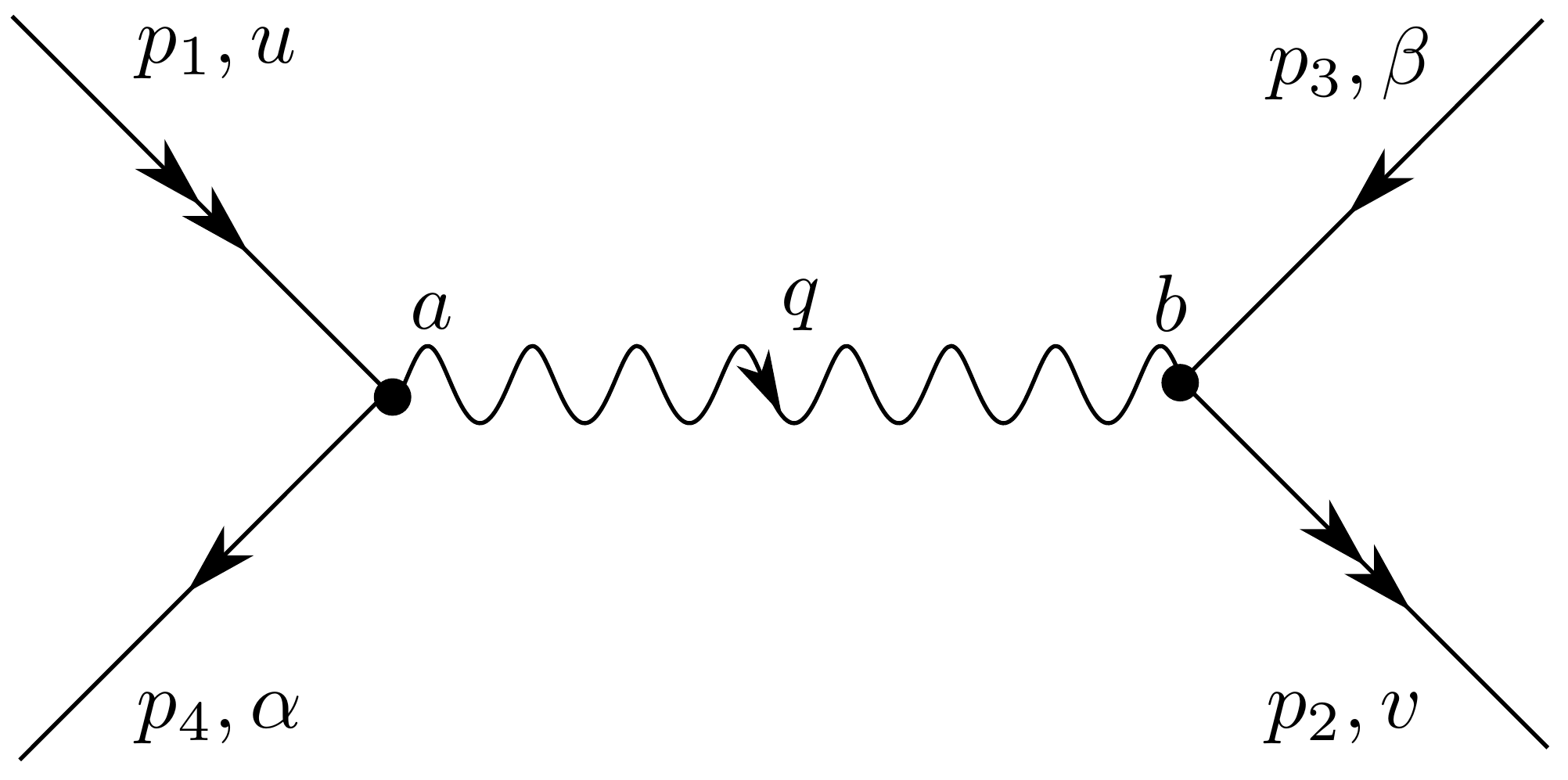}
  \caption{The vertex generated by the action $S_{\Delta c}$, see Eq.~\eqref{a-SDc}, describing interaction between a molecule (denoted with a solid line with a double
    arrow) and impurities (denoted with a solid line with a single arrow). Each line carries the momentum variable and the Keldysh index. The interaction mediated by
    fermions is denoted by the wavy line, which is labeled by a momentum variable $q$ and two Keldysh indices (together they determine $G_{f,ab}(q)$). To the diagram we
    assign the expression that is shown in Eq.~\eqref{a-vert-val}.}
\label{fig:vertex}
\end{figure}

In Fig.~\ref{fig:vertex}, we display the vertex generated by the action $S_{\Delta c}$ defined by Eq.~\eqref{a-SDc}. Each line carries a momentum and a Keldysh index. A
solid line with an arrow denotes the impurity field, while double arrows represent the molecular fields. When forming diagrams, two lines are connected to form a GF that,
depending on the carried Keldysh indices, is taken as the retarded/advanced or Keldysh GF, as shown in Tab.~\ref{table}. In the figure, the wavy line stands for the
fermion GF. It carries an index at both ends, denoted by $a$ and $b$, and a momentum $q$. The diagram is thus translated into the expression
\begin{equation}
\label{a-vert-val}
  -\frac12 \frac{1}{V} \gamma^{\alpha}_{u,a} G_{f,ab}(q) \gamma^{\beta}_{b,v},
\end{equation}
where $q = p_1 - p_4 = p_2 - p_3$. 

%
%

From the vertex shown in Fig.~\ref{fig:vertex} we can evaluate the self-energies. To calculate the molecular self-energy, we connect the external impurity lines, i.e., we
set $p_3=p_4$, and introduce $G_{0,c,\alpha\beta}(p_3)$ for that line. In this way, we arrive at the self-energy shown in Fig.~\ref{fig:diag_xD}b. When the sum over
Keldysh indices is evaluated, we obtain a sum of diagrams in which retarded/advanced and Keldysh GFs appear, see Fig.~\ref{fig:diag_xD}b.  Analogously, by joining the
molecular line, i.e., by setting $p_1=p_2$ and introducing $G_{\Delta,uv}(p_1)$, we obtain the diagram from Fig.~\ref{fig:diag_xD}a. Taking the sum over the Keldysh
indices we arrive at the sum of diagrams, see Fig.~\ref{fig:diag_xD}a. However, in the resummation scheme employed in this work, in the calculation of the impurity
self-energy the dressed molecular GF is used instead of the bare one.

\subsection{Dyson equation}
\label{app-dyson}

The action of the system, after tracing out the fermions and introducing the molecular channel consists of three parts
\begin{equation}
  S = S_c + S_\Delta + S_{\Delta c}.
\end{equation}
The first two terms on the right-hand side generate the impurity and molecular bare Green functions, respectively. The third term is responsible 
for the conversion of a fermion and an impurity in an impurity-fermion molecule. 

The evaluation of the dressed GFs, which are given by average over the field weighted by $e^{iS}$, can be understood as averaging over $e^{i S_c + i S_\Delta}$ and an expansion of $e^{iS_{\Delta c}}$ in  powers of the interaction. The
usual diagrammatic formulation of the theory is then recovered when we assign lines to the GFs as shown in Tab.~\ref{table}.

As a result of the summation of the diagrams, we obtain Dyson equations for the GFs:
\begin{subequations}
  \begin{eqnarray}
    G_c &=& G_{0,c} + G_{0,c} \cdot \Sigma_c \cdot G_c, \label{DEc}\\
    G_\Delta &=& G_{0,\Delta} + G_{0,\Delta} \cdot \Sigma_\Delta \cdot G_\Delta, \label{DED}\\
  \end{eqnarray}
\end{subequations}
where by $\cdot$ we denote the multiplication of the Keldysh matrices. Here, the self-energies $\Sigma_c$ and $\Sigma_\Delta$ are matrices generated by irreducible diagrams for
the impurity and molecular GFs, respectively. They have a causal structure similar to the inverse GFs~\cite{kamenev2011field}:
\begin{equation}
\label{a-SED}
  \Sigma_\Delta = 
  \left( 
  \begin{array}{cc}
    \Sigma_\Delta^R & \Sigma_\Delta^K \\
    0 & \Sigma_\Delta^A 
  \end{array} \right)
\end{equation}
for the molecule, and
\begin{equation}
\label{a-SEx}
  \Sigma_c = 
  \left( 
  \begin{array}{cc}
    0 & \Sigma_c^A \\
    \Sigma_c^R & \Sigma_c^K
  \end{array} \right)
\end{equation}
for the impurity.

The self-energy of the molecule is generated by the diagram shown in Fig.~\ref{fig:diag_xD}b. Employing the vertex from the previous appendix, we arrive at
\begin{eqnarray}
\label{a-SED2}
  \Sigma_{\Delta,ab}(p) &=& \frac{i}{2V}\sum_{p'} \gamma^\alpha_{aa'}\gamma_{bb'}^\beta \times \nonumber \\
  && G_{f,a'b'}(p-p')G_{0,c,\alpha\beta}(p').
\end{eqnarray}
The retarded self-energy is thus given by $\Sigma_\Delta^R = [\Sigma_\Delta]_{11}$, which leads to Eq.~\eqref{SED}.

Analogously, the diagram from Fig.~\ref{fig:diag_xD} generates the self-energy for the impurity. Here, we take the dressed GFs for the molecule with self-energy given
by~\eqref{a-SED2}. As a result, we obtain:
\begin{equation}
\label{a-SEx2}
  \Sigma_x^{\alpha\beta}(p) = \frac{-i}{2V}\sum_{p'} \tr{ \gamma^\beta\cdot G_f(p') \cdot \gamma^\alpha \cdot G_\Delta(p+p')},
\end{equation}
where the trace acts in the Keldysh indices.
Finally, using Eq.~\eqref{a-SEx} we find $\Sigma_x^R = [\Sigma_x]_{q,cl}$, from which follows Eq.~\eqref{SEx}.


%
%

%

\end{document}